\documentclass [twocolumn]{aastex62}
\usepackage[top=1in, bottom=1in, left=1in, right=1in]{geometry}
\usepackage{hyperref} 
\usepackage{amssymb, amsmath, mathtools}
\usepackage{natbib}
\usepackage{enumerate}
\usepackage{graphicx}

\usepackage{latexsym}
\usepackage{color}
\usepackage{graphicx}
\usepackage{epstopdf}
\usepackage{threeparttable}
\usepackage[T1]{fontenc}
\usepackage{tabularx}
\usepackage{multirow}
\usepackage{colortbl}

\newcommand\degree{\degr}
\newcommand\degrees\degree

\newcommand\jwst{\em JWST}

\DeclareSymbolFont{UPM}{U}{eur}{m}{n}
\DeclareMathSymbol{\umu}{0}{UPM}{"16}
\let\oldumu=\umu
\renewcommand\umu{\ifmmode\oldumu\else\math{\oldumu}\fi}
\newcommand\micro{\umu}
\newcommand\microns \micron

\let\oldsim=\sim
\renewcommand\sim{\ifmmode\oldsim\else\math{\oldsim}\fi}
\let\oldpm=\pm
\renewcommand\pm{\ifmmode\oldpm\else\math{\oldpm}\fi}
\newcommand\by{\ifmmode\times\else\math{\times}\fi}

\newbox{\wdbox}
\renewcommand\c{\setbox\wdbox=\hbox{,}\hspace{\wd\wdbox}}
\renewcommand\i{\setbox\wdbox=\hbox{i}\hspace{\wd\wdbox}}

\newcount\timect
\newcount\hourct
\newcount\minct
\newcommand\now{\timect=\time \divide\timect by 60
         \hourct=\timect \multiply\hourct by 60
         \minct=\time \advance\minct by -\hourct
         \number\timect:\ifnum \minct < 10 0\fi\number\minct}

\catcode`@=11

\newcommand\comment[1]{}
\newcommand\commenton{\catcode`\%=14}
\newcommand\commentoff{\catcode`\%=12}

\renewcommand\math[1]{$#1$}
\newcommand\mathshifton{\catcode`\$=3}
\newcommand\mathshiftoff{\catcode`\$=12}

\comment{the backslash is necessary}

\comment{alignment tab}

\let\atab=&
\newcommand\atabon{\catcode`\&=4}
\newcommand\ataboff{\catcode`\&=12}

\let\oldmsp=\sp
\let\oldmsb=\sb
\def\sp#1{\ifmmode
           \oldmsp{#1}%
         \else\strut\raise.85ex\hbox{\scriptsize #1}\fi}
\def\sb#1{\ifmmode
           \oldmsb{#1}%
         \else\strut\raise-.54ex\hbox{\scriptsize #1}\fi}
\newbox\@sp
\newbox\@sb
\def\sbp#1#2{\ifmmode%
           \oldmsb{#1}\oldmsp{#2}%
         \else
           \setbox\@sb=\hbox{\sb{#1}}%
           \setbox\@sp=\hbox{\sp{#2}}%
           \rlap{\copy\@sb}\copy\@sp
           \ifdim \wd\@sb >\wd\@sp
             \hskip -\wd\@sp \hskip \wd\@sb
           \fi
        \fi}
\def\msp#1{\ifmmode
           \oldmsp{#1}
         \else \math{\oldmsp{#1}}\fi}
\def\msb#1{\ifmmode
           \oldmsb{#1}
         \else \math{\oldmsb{#1}}\fi}
\def\supon{\catcode`\^=7}
\def\supoff{\catcode`\^=12}
\def\subon{\catcode`\_=8}
\def\suboff{\catcode`\_=12}
\def\supsubon{\supon \subon}
\def\supsuboff{\supoff \suboff}

\newcommand\actcharon{\catcode`\~=13}
\newcommand\actcharoff{\catcode`\~=12}

\newcommand\paramon{\catcode`\#=6}
\newcommand\paramoff{\catcode`\#=12}

\comment{And now to turn us totally on and off...}

\newcommand\reservedcharson{\commenton \mathshifton \atabon \supsubon \actcharon
	\paramon}

\newcommand\reservedcharsoff{\commentoff \mathshiftoff \ataboff
	\supsuboff \actcharoff \paramoff}

\catcode`@=12
\reservedcharsoff

\reservedcharson

\comment{ Must have ONLY ONE of these... trust these macros, they work

}

\newcommand{\squishlist}{
 \begin{list}{$\bullet$}
  { \setlength{\itemsep}{0pt}
     \setlength{\parsep}{0pt}
     \setlength{\topsep}{0pt}
     \setlength{\partopsep}{0pt}
     \setlength{\leftmargin}{2.0em}
     \setlength{\labelwidth}{1.5em}
     \setlength{\labelsep}{0.5em} } }

\newcommand{\squishlisttwo}{
 \begin{list}{$\bullet$}
  { \setlength{\itemsep}{1pt}
     \setlength{\parsep}{3pt}
     \setlength{\topsep}{3pt}
     \setlength{\partopsep}{0pt}
     \setlength{\leftmargin}{2.0em}
     \setlength{\labelwidth}{1.5em}
     \setlength{\labelsep}{0.5em} } }

\newcommand{\squishend}{
  \end{list}  }

\reservedcharson
\actcharon
\graphicspath{{./}{Figures/}}

\shorttitle{Sample article}
\shortauthors{Schwarz et al.}

\begin{document}

\title{On the Utility of Transmission Color Analysis I: Differentiating Super-Earths and Sub-Neptunes}
\markright{On the Utility of Transmission Color Analysis for Differentiating Super-Earths and Sub-Neptunes, K.S. Sotzen et al}

\correspondingauthor{Kristin Showalter Sotzen}
\email{kristin.sotzen@jhuapl.edu, kshowal3@jhu.edu}

\author[0000-0001-7393-2368]{Kristin S. Sotzen}
\affiliation{ JHU Applied Physics Laboratory, 11100 Johns Hopkins Rd, Laurel, MD 20723, USA}
\affil{Johns Hopkins University, 3400 N. Charles St, Baltimore, MD 21218, USA}

\author{Kevin B. Stevenson}
\affiliation{ JHU Applied Physics Laboratory, 11100 Johns Hopkins Rd, Laurel, MD 20723, USA}

\author[0000-0002-2739-1465]{Erin M. May}
\affiliation{ JHU Applied Physics Laboratory, 11100 Johns Hopkins Rd, Laurel, MD 20723, USA}

\author{Natasha E. Batalha}
\affiliation{ NASA Ames Research Center, MS 245-3, Moffett Field, CA 94035, USA}
\author{Noam R. Izenberg}
\affiliation{ JHU Applied Physics Laboratory, 11100 Johns Hopkins Rd, Laurel, MD 20723, USA}

\author{Sarah M. H{\"o}rst}
\affil{Johns Hopkins University, 3400 N. Charles St, Baltimore, MD 21218, USA}
\affiliation{Space Telescope Science Institute, 3700 San Martin Dr, Baltimore, MD 21218, USA}

\author{Calley L. Tinsman}
\affiliation{ JHU Applied Physics Laboratory, 11100 Johns Hopkins Rd, Laurel, MD 20723, USA}

\author{Carey M. Lisse}
\affiliation{ JHU Applied Physics Laboratory, 11100 Johns Hopkins Rd, Laurel, MD 20723, USA}

\author{Nikole K. Lewis}
\affiliation{Department of Astronomy and Carl Sagan Institute, Cornell University, 122 Sciences Drive, Ithaca, NY 14853, USA }

\author{Jayesh M. Goyal}
\affiliation{Department of Astronomy and Carl Sagan Institute, Cornell University, 122 Sciences Drive, Ithaca, NY 14853, USA }
\affiliation{National Institute of Science Education and Research (NISER), Jatni, Khurda-752050, Odisha, India}

\author{Joseph J. Linden}
\affiliation{ JHU Applied Physics Laboratory, 11100 Johns Hopkins Rd, Laurel, MD 20723, USA}

\author{Kathleen E. Mandt}
\affiliation{ JHU Applied Physics Laboratory, 11100 Johns Hopkins Rd, Laurel, MD 20723, USA}

\received{February 2021}
\revised{May 2021}
\accepted{June 2021}

\begin{abstract}

The majority of exoplanets found to date have been discovered via the transit method, and transmission spectroscopy represents the primary method of studying these distant worlds. Currently, in-depth atmospheric characterization of transiting exoplanets entails the use of spectrographs on large telescopes, requiring significant observing time to study each planet. Previous studies have demonstrated trends for solar system worlds using color-color photometry of reflectance spectra, as well as trends within transmission spectra for hot Jupiters. Building on these concepts, we have investigated the use of transmission color photometric analysis for efficient, coarse categorization of exoplanets and for assessing the nature of these worlds, with a focus on resolving the bulk composition degeneracy to aid in discriminating super-Earths and sub-Neptunes. 

We present our methodology and first results, including spectrum models, model comparison frameworks, and wave band selection criteria. We present our results for different transmission ``color'' metrics, filter selection methods, and numbers of filters. Assuming noise-free spectra of isothermal atmospheres in chemical equilibrium, with our pipeline, we are able to constrain atmospheric mean molecular weight in order to distinguish between super-Earth and sub-Neptune atmospheres with >90{$\%$} overall accuracy using as few as two specific low-resolution filter combinations. We also found that increasing the number of filters does not substantially impact this performance. This method could allow for broad characterization of large numbers of planets much more efficiently than current methods permit, enabling population and system-level studies. Additionally, data collected via this method could inform follow-up observing time by large telescopes for more detailed studies of worlds of interest.

\end{abstract}



\section{Introduction}
\label{sec:intro}

Identifying habitable planets around other stars is one of NASA’s greatest long-term goals. Thanks to the \emph{Kepler} and \emph{Transiting Exoplanet Survey Satellite (TESS)} space telescopes \citep{Borucki2010,Ricker2015}, transiting planets dominate the current population of confirmed exoplanet discoveries, and transmission and emission spectra represent the primary method of studying these distant worlds. For the foreseeable future, atmospheric characterization with telescopes like the \emph{Hubble Space Telescope (HST)} and the \emph{James Webb Space Telescope} ({\jwst}) will provide windows into understanding the nature of exoplanets. However, there are strong technological and observing time limitations that will constrain our ability to collect detailed spectra on large numbers of exoplanets using these high-demand telescopes.

Current methods of characterizing transiting exoplanets entail the use of multiple spectrographs -- sometimes on multiple telescopes like \emph{HST}, the \emph{Spitzer Space Telescope}, and ground-based observatories -- to achieve broad wavelength coverage (\emph{e.g.,} \cite{Stevenson2014a, Sing2016, Wakeford2017b, Sotzen2020, Alam2021}). This process typically requires significant observing time to study each planet. With thousands of confirmed planets, we are approaching a stage where survey observations and statistical studies will afford a more contextual and holistic understanding of aggregate planetary system architectures and evolution than can be achieved from examining a few dozen atmospheres in detail.

Solar system studies have highlighted the great promise of characterizing planetary bodies using color-color photometry.  \cite{Traub2003} identified trends in solar system bodies using broad blue, green, and red filters in the visible and near-infrared. \cite{Crow2011} showed that color-color reflectance ratios – i.e., comparison of reflected fluxes for specific filters – can be used to broadly categorize solar system bodies, with Earth occupying a unique position in reflectance color space for particular wave bands.  \cite{Krissansen-Totton2016} later demonstrated that specific visible filters can be used to uniquely identify an Earth-like spectrum in color-color space, and that, if the observational noise is dominated by dark current, the integration time required to identify Earth’s unique colors is $\sim$20 times shorter than the integration time required to obtain a moderate resolution (R $\sim$ 70) spectrum.

Spectroscopic and photometric trends have also been identified for observed and simulated planets beyond the Solar System. \cite{Sing2016} showed trends in hot Jupiter water abundances as a function of blue-optical vs NIR/MIR (near-infrared/mid-infrared) altitude differences, and \cite{Stevenson2016b} demonstrated trends in hot Jupiter water abundances as a function of temperature and gravity. \cite{Gao2020} found that, for warm extrasolar giant planets, variations in cloudiness -- indicated by the amplitude of the 1.4-$\mu$m H$_2$O feature -- are dictated primarily by temperature due to the formation of silicate clouds at T\textsubscript{eq}>950 K and hydorcarbon hazes at T\textsubscript{eq}<950 K. Additionally, \cite{Baxter2020} used \emph{Spitzer} and \emph{HST} Wide Field Camera 3 (WFC3) to compute an emission ``color'' difference brightness for a set of observed hot and ultrahot extrasolar giants, for which they identified a trend from stronger water absorption features toward a blackbody with increasing equilibrium temperature. Moreover, \cite{Crossfield2017} found a positive correlation between the amplitude of the 1.4-$\mu$m H$_2$O feature and the temperature and atmospheric H/He abundance for 6 Neptune-sized planets with radii of 2--6 R$_\oplus$ and temperatures of 500-1000 K.  

\cite{Batalha2018} also showed that it is possible to classify giant planets in reflectance color-color (i.e., color ratios) space using WFIRST-like filters for planets that do not have significant cloud coverage. \cite{Grenfell2020} went on to investigate the utility of transmission depth differences for the filters of the PLAnetary Transits and Oscilllations of stars (PLATO) mission, showing that basic atmospheric types (primary and water-dominated) and the presence of submicron hazes could be distinguished for some planets.

Reflectance color-color trends ensue from correlations between the relative strengths of different reflectance features and characteristics of a planet or world (\emph{e.g.,} the presence and type of atmosphere; \cite{Traub2003, Crow2011, Triaud2014a, Triaud2014b, Batalha2018, Dransfield2020, Grenfell2020, Melville2020}).  For example, \cite{Crow2011} found that the solar system worlds, and Earth in particular, can be categorized based on the presence or predominance of Rayleigh scattering in the ultraviolet and blue wavelengths. In principle, this technique can be extrapolated to transmission spectra, where the relative strengths of different molecular features, and/or the variation of the peak/trough relationship or slope of a particular molecular feature, are correlated with atmospheric properties such as temperature or mean molecular weight (MMW). Here we have investigated the potential for the color analysis method for use with transmission spectra for coarsely characterizing exoplanet atmospheres using low-resolution spectra, with the goal of identifying an optimal filter subset or wavelength range for distinguishing super-Earths from sub-Neptunes. 

``Super-Earth'' generally refers to a rocky planet with a radius between 1.25 and approximately 1.75 Earth radii (1.25$R_\oplus$ < R$_P$ < $\sim$1.75$R_\oplus$) with a secondary atmosphere lacking H and He, while ``sub-Neptune'' generally refers to a planet with a radius larger than approximately 1.75$\times$ Earth's radius but smaller than Neptune's ($\sim$1.75$R_\oplus$ < R$_P$ < 3$R_\oplus$) with a substantial gaseous envelope comprising a primary atmosphere with a significant fraction of H and He \citep{Rogers2010, Batalha2013, Benneke2013, Lopez2014, Marcy2014b, Weiss2014, Rogers2015, Fulton2017}. Given the wealth of confirmed transiting exoplanets, a color analysis method optimized for transmission spectra has the potential to facilitate rapid characterization of these worlds and to enable statistical studies of exoplanet atmospheres, as well as investigations into broader questions pertaining to the Radius Gap \citep{Fulton2017} and the super-Earth/sub-Neptune degeneracy \citep{Rogers2010, Rogers2015}. Transmission color photometric analysis also has applications in determining potential biomarkers on Earth-sized worlds (\cite{Lisse2020} and references therein). 

For this initial study, we have focused on super-Earth and sub-Neptune sized exoplanets, with the goal of resolving the bulk composition degeneracy between these classes of objects. Based on the \emph{Kepler} data, these two categories constitute the majority of exoplanet discoveries to date \citep{Batalha2013, Fulton2018, Berger2020}, so a more complete picture of the demographics of these two classes of planets will facilitate a better understanding of the processes that shape them and the implications for super-Earth habitability.

Super-Earths and sub-Neptunes can have similar radii, and transit observations of these planets require follow-up with radial-velocity (RV) observations to determine their masses and assess their bulk densities and likely compositions.  However, even precise mass and radius measurements can lead to ambiguous bulk densities for planets that fall in the super-Earth\slash sub-Neptune transition region \citep{Fortney2007, Seager2007}, necessitating observations of such a planet's atmosphere to assess its planetary classification.  As shown in \citet{Yu2021}, certain atmospheric species may be used to distinguish between worlds with surfaces at different depths; however, these measurements will be difficult to make with current observing techniques.  The goal of this investigation was to facilitate quickly and efficiently distinguishing between these classes of exoplanets by identifying correlations between transmission colors and\slash or color ratios and planetary parameters (\textit{e.g.,} mass, temperature, and metallicity), with a focus on estimating atmospheric mean molecular weight as an indicator of planet category (\textit{i.e.,} super-Earth or sub-Neptune). Our primary goal was to determine whether this methodology can work in principle, so this initial investigation was conducted using noise-free spectra. Extending the investigation to include noise was out of scope for this study, and will be explored in subsequent analyses.

Insights from our results can be used to inform observation planning for transiting exoplanets by telescopes such as {\jwst} and to guide future instrument and mission concept development.

In Section \ref{sec:simspec}, we describe the parameters and tools used to generate our database of simulated spectra, and in Section \ref{sec:filters}, we review the wavelength limits and resolution of our synthetic color analysis.  In Sections \ref{sec:metric} and \ref{sec:results}, we discuss our transmission color-color metric, analysis methods, and results, and we address the implications of our findings.  Section \ref{sec:future} outlines our planned future work, and Section \ref{sec:conclusion} reviews our conclusions.

\section{Simulated Spectra \& Molecular Weight Groups} \label{sec:simspec}

For this investigation, we used Machine Learning with a grid framework of model transmission spectra, applying the Machine Learning technique across a wide planet parameter space to capture the variation in transmission spectra over the expected and applicable range of planetary and atmospheric parameters. Therefore, we used an expansion of the library of forward model transmission spectra and corresponding chemical equilibrium abundances presented by \citet{Goyal2019}. This set of simulated exoplanet transmission spectra were generated for general community use using the ATMO 1D-2D radiative-convective equilibrium atmosphere code \citep{Amundsen2014, Tremblin2015, Drummond2016, Goyal2018}. Although these spectra are nominally computed for a Jupiter radius planet (1 R\textsubscript{Jup} at 1 millibar pressure) around a Solar radius star (1 R\textsubscript{Sun}), as shown in \cite{Goyal2019}, they can be scaled and applied to a wide variety of parameters (temperature, metallicity, gravity, etc) to generate a spectrum for a customized planet. 

Since we targeted the question of differentiating between super-Earths and sub-Neptunes in the Radius Gap, we focused this study on planets with radius R = 1.75\textsubscript{Earth} \citep{Fulton2017}. Planets of this size are expected to have thick atmospheres \citep{Lopez2014}, and dynamical surface effects are not expected to impact the atmosphere at the 1 mbar (0.001 bar) pressure level \citep{May2020}, which approximates the region of the atmosphere probed with transmission spectra \citep{LecavelierDesEtangs2008}. While we expect planets of these sizes generally to have thick atmospheres, it is theoretically possible for super-Earths to have thin or tenuous atmospheres, and our 1 mbar assumption approaches the pressures for these thinner atmospheres (\emph{e.g.,} Mars).

As the focus of this effort was to differentiate between expected atmospheric categories based on transmission spectra simulated for a range of atmospheric compositions at the 1 mbar pressure level, the exact chemical and circulation processes that result in these 1 mbar compositions do not help to address the question at hand. Given the aforementioned assumptions and the paucity of modeling and observational constraints on the  compositions of sub-Neptune and super-Earth atmospheres \citep{Moran2020}, we consider the scaling of the ATMO spectra to super-Earth and sub-Neptune parameters to be a valid assumption for this study. 

For the library of transmission spectra presented in this work, we used isothermal $P$-$T$ profiles consistent with equilibrium chemistry. The isothermal $P$-$T$ profiles extend from 10\textsuperscript{-6} bar at the top of the atmosphere to 10 bar at the bottom, with a radius for the simulated baseline planet that is 1 R\textsubscript{Jup} defined at the 1 mbar pressure level. We used H$_2$-H$_2$ and H$_2$-He collision induced absorption (CIA) opacities, as well as opacities from 18 other absorption species (H$_2$O, CO$_2$, CO, CH$_4$, NH$_3$, Na, K, Li, Rb, Cs, TiO, VO, FeH, PH$_3$, HCN, C$_{2}$H$_{2}$, H$_{2}$S and SO$_{2}$; see \citet{Goyal2018} for additional details).  The spectral library has a R$\sim$5000 at 0.2 {\micro m} while decreasing to R$\sim$100 at 10 {\micro m}.  We acknowledge that ATMO's current treatment of cloud particles and haze parameters is simplistic and does not represent the entire diversity of aerosols that could be present in planetary atmospheres.  However, given the aforementioned lack of constraints on these parameters for exoplanet atmospheres \citep{Moran2020}, we applied the most straightforward approach commonly used in the literature to interpret current observations of exoplanet atmospheres \citep{Sing2016}.

ATMO provides spectra for two different approaches to calculating chemical equilibrium abundances of the species considered: (1) the local condensation approach, which assumes that condensates only deplete the material in the specific layer of the atmosphere where they form, but not for higher levels of the atmosphere, and (2)  the rainout condensation approach, which assumes that condensation depletes material from the first layer of formation and for all layers above. We only used the rainout condensation models, as these include extinction of the condensate species above the initial altitude of condensation \citep{Goyal2018}, which we consider to be more physically accurate than the local condensation models. A simplified treatment of clouds and hazes is applied in the \cite{Goyal2019} spectral library that increases the flexibility and utility of the library while matching the fidelity of current observational constraints on exoplanet aerosols. Haze is treated as small scattering aerosol particles and implemented as a parameterized enhanced multi-gas Rayleigh scattering, while clouds are treated as large particles with gray opacity \citep[see][for details]{Goyal2019}. 
\cite{Goyal2019} computed models over a range of 22 planetary equilibrium temperatures (400, 600, 700, 800, 900, 1000, 1100, 1200, 1300, 1400, 1500, 1600, 1700, 1800, 1900, 2000, 2100, 2200, 2300, 2400, 2500, 2600 K), four planetary surface gravities (5, 10, 20, 50; all listed in ms\textsuperscript{-2} and equivalent to 0.5, 1.0, 2.0, and 5.0 g), five atmospheric metallicities (1, 10, 50, 100, 200; all in \emph{x solar}), four C/O ratios (0.35, 0.56, 0.7, 1.0), four scattering haze parameters (1, 10, 100, 1100x standard Rayleigh-scattering) and four uniform cloud particle cross-sections (0, 3E-4, 1E-3, and 5E-3 cm$^2$/g).   

For this work, we expanded this library to include additional cooler temperatures and high metallicities -- specifically, temperature values of 300 and 500\,K, and metallicity values of 1000, 1500, 3000, and 10,000 times the solar value. These higher-metallicity models allow for the investigation of higher mean molecular weight (\emph{e.g.,} H$_2$O- and CO$_2$-rich) atmospheres in addition to the H- and He-rich atmospheres provided by the lower-metallicity models.  This provided a full range of representative atmospheric mean molecular weights, which was the parameter by which we assigned our atmospheric models as either super-Earths or sub-Neptunes. 

We applied our analysis technique to the models for all of the C/O ratios, haze parameters, and cloud parameters.  Based on the thermal escape parameter described in \cite{Konatham2020}, we sought to limit our temperature parameter to values that would be realistic for planets in our mass range of interest.  Given the recent discovery of the ultrahot Neptune-mass exoplanet LTT 9779 b \citep{Jenkins2020} in the `hot Neptune desert' \citep{Mazeh2016} and the detection of an atmosphere around the super-Earth 55 Cancri e \citep{Demory2016,Angelo2017}, both with equilibrium temperatures $\geq$2000 K, we limited our analysis to models with temperatures $\leq$2200 K in an effort to capture the theoretical limits of atmospheric retention while acknowledging relevant observational data.

In this study, we sought to define observational discriminators between super-Earths and sub-Neptunes; therefore, this effort focused on worlds with radius R = 1.75 R\textsubscript{Earth}, as this is a recommended division between these planet classes \citep{Lopez2014}.  Given the Signal-to-Noise Ratio (SNR) considerations associated with the atmospheres of these smaller planets, we assumed a M-dwarf host star with R = 0.5 R\textsubscript{Sun}.  M dwarfs constitute the majority ($\sim$75\%) of stars in the solar neighborhood, and planets orbiting M dwarfs are of increasing interest due to the favorable signal sizes of their atmospheric molecular features, compared to planets around brighter stars.  It is worth noting that the \cite{Goyal2019} models assume a stellar radiance distribution similar to the Sun, which may result in slightly different chemistry from that of a smaller, cooler star.  However, we assess that this assumption is acceptable for this initial study based on the reasoning applied to our isothermal and chemical equilibrium assumptions.

Here we focused on the 10, 16, and 20 ms\textsuperscript{-2} cases, as they best span the gravities in the current super-Earth/sub-Neptune populations \citep{Wu2019,Gupta2020}.  The \cite{Goyal2019} grid of transmission models includes the surface gravities listed previously, and we used the scaling script provided with the \cite{Goyal2019} grid to estimate the spectra for the 16 ms\textsuperscript{-2} surface gravity case by scaling down from the corresponding 20 ms\textsuperscript{-2} spectra.  Our interest lay in approximating the super-Earth/sub-Neptune mass range for the R = 1.75 M\textsubscript{Earth}, and these gravities correspond to planets with masses of \sim 3.1, 5.0, and 6.1 M\textsubscript{Earth}.  As Neptune has an atmospheric metallicity of \sim 100x solar \citep{Karkoschka2011, Wakeford2020}, we limited our metallicity selection to $\geq$200x solar to capture the range of expected atmospheric mean molecular weights for super-Earths and sub-Neptunes.  The \cite{Goyal2019} grid includes a chemical profile for each simulated transmission spectrum, and we computed the mean molecular weight for each spectrum based on the molecular abundances at 1 millibar; this computation would not be significantly affected by the rainout chemistry, as minor species would be more likely to condense out below 1 millibar.  We used the aforementioned scaling script to scale all of the spectra for a R = 0.5 R\textsubscript{Sun} star.

The combination of temperatures $\leq$2200 K, three surface gravities, five metallicities, and four each for C/O ratio, haze parameters, and cloud parameters yields a total of \sim 19,000 model spectra over which to explore statistical correlations, which enables robust statistical analysis and is comparable to other Machine Learning studies \citep[e.g.,][]{Batalha2018}. Table \ref{tab:modelParams} provides a summary of our model parameters.

\begin{deluxetable}{c|c}
\tablecaption{Model Parameters for Transmission Color Analysis \label{tab:modelParams}}
\tablehead{\colhead{Model Parameter} & \colhead{Value} \\
}
\startdata
Star Radius ($R_\odot$) & 0.5\\
Planet Radius ($R_\oplus$) & 1.75\\
Planet g (ms\textsuperscript{-2}) & 10, 16, 20\\
$T_P$ (K) & 300 -- 2200, steps of 100\\
Metallicity (\emph{$\times$ solar}) & 200, 1k, 1.5k, 3k, 10k\\
C/O Ratio & 0.35, 0.56, 0.70, 1.00\\
Cloud Particle & 0, 3E-4, 1E-3, 5E-3\\
Cross-Section (cm\textsuperscript{2}/g) &\\
Haze Parameter & 1, 10, 100, 1100\\
($\times$ standard Rayleigh scattering) &\\
\enddata
\end{deluxetable}

\begin{figure}[t]
\includegraphics[width=1.0\linewidth]{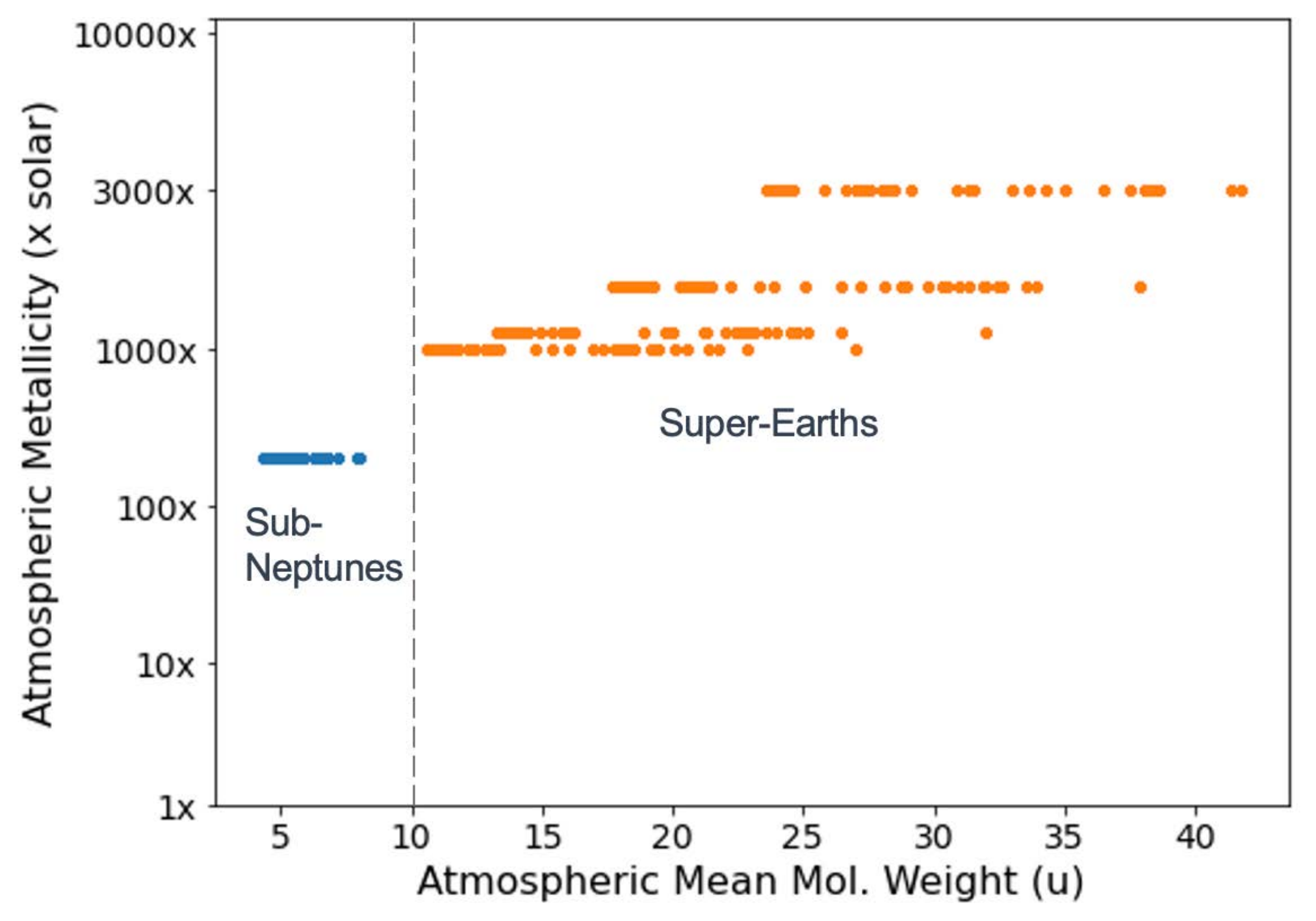}
\caption{\label{fig:mmw_vs_met}\small
Mean molecular weight of our simulated transmission spectra as a function of metallicity. Lower metallicities produce spectra with more consistent MMWs, whereas spectra with higher metallicities show more scatter in the MMW values. These data show a visible division at 10 u, which we used as our threshold for planet categorization -- sub-Neptunes are assumed to have MMW $\leq$10 u, and super-Earths are assumed to have MMW $>$10 u \citep{Benneke2013}.}
\end{figure}

\begin{figure}[h]
\includegraphics[width=1.0\linewidth]{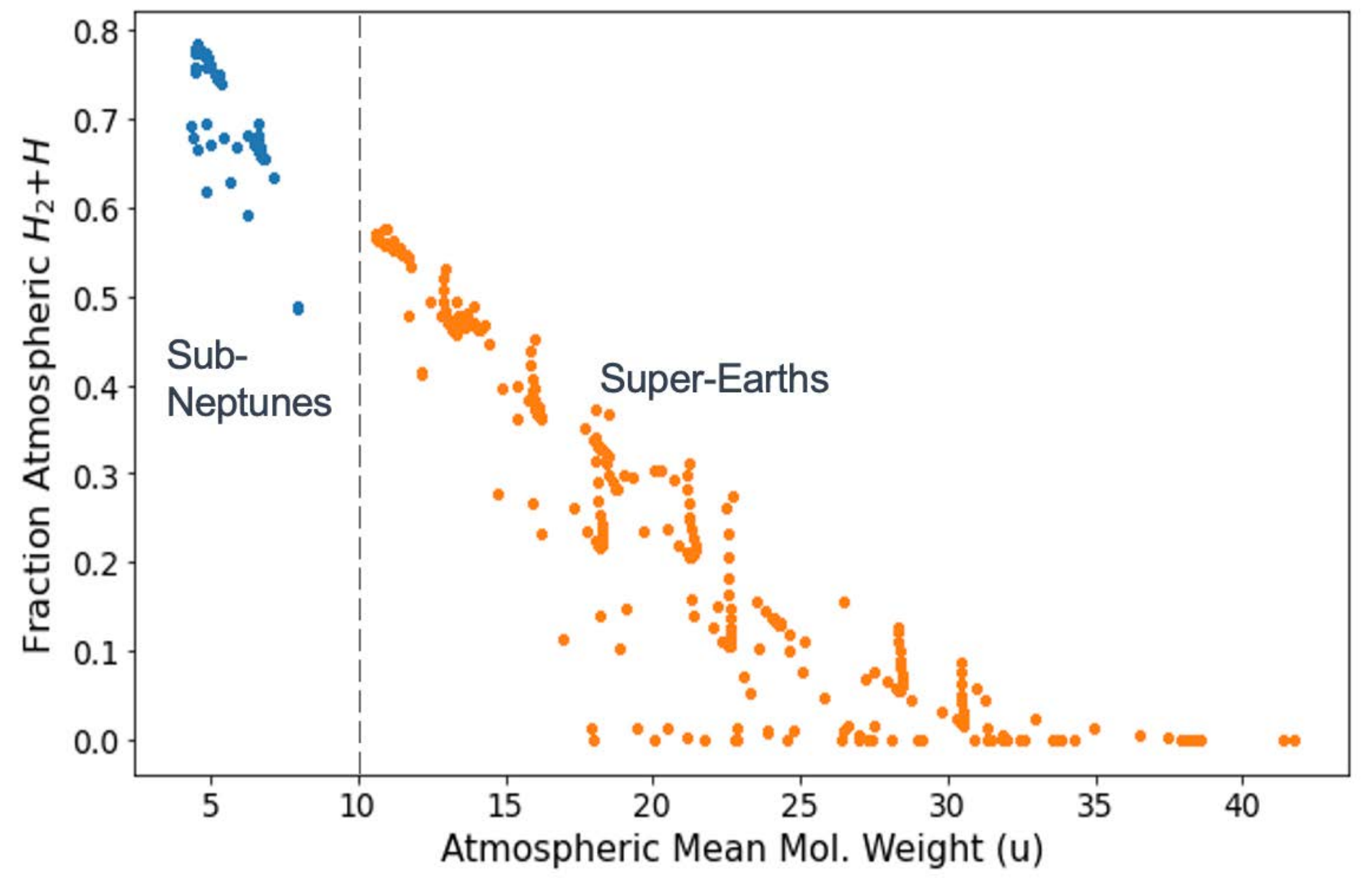}
\caption{\label{fig:h_vs_mmw}\small
Hydrogen abundance (H$_2$+H) as a function of atmospheric mean molecular weight for our simulated transmission spectra with temperatures $\leq$2200 K. There is a noticeable division spanning 7.14 u and 10.62 u, corresponding to hydrogen abundances of 63.3\% and 56.5\%, respectively. A small cluster at \sim 48.5\% hydrogen abundance corresponds to a MMW of 7.96 u and is associated with the T = 400 \& 500 K spectra with C/O ratio of 0.35. This sub-solar C/O ratio provides for Oxygen in the atmosphere and favoring formation of CO$_2$, increasing the atmospheric MMW for this cluster. }
\end{figure}

\begin{figure}[t]
\includegraphics[width=1.0\linewidth]{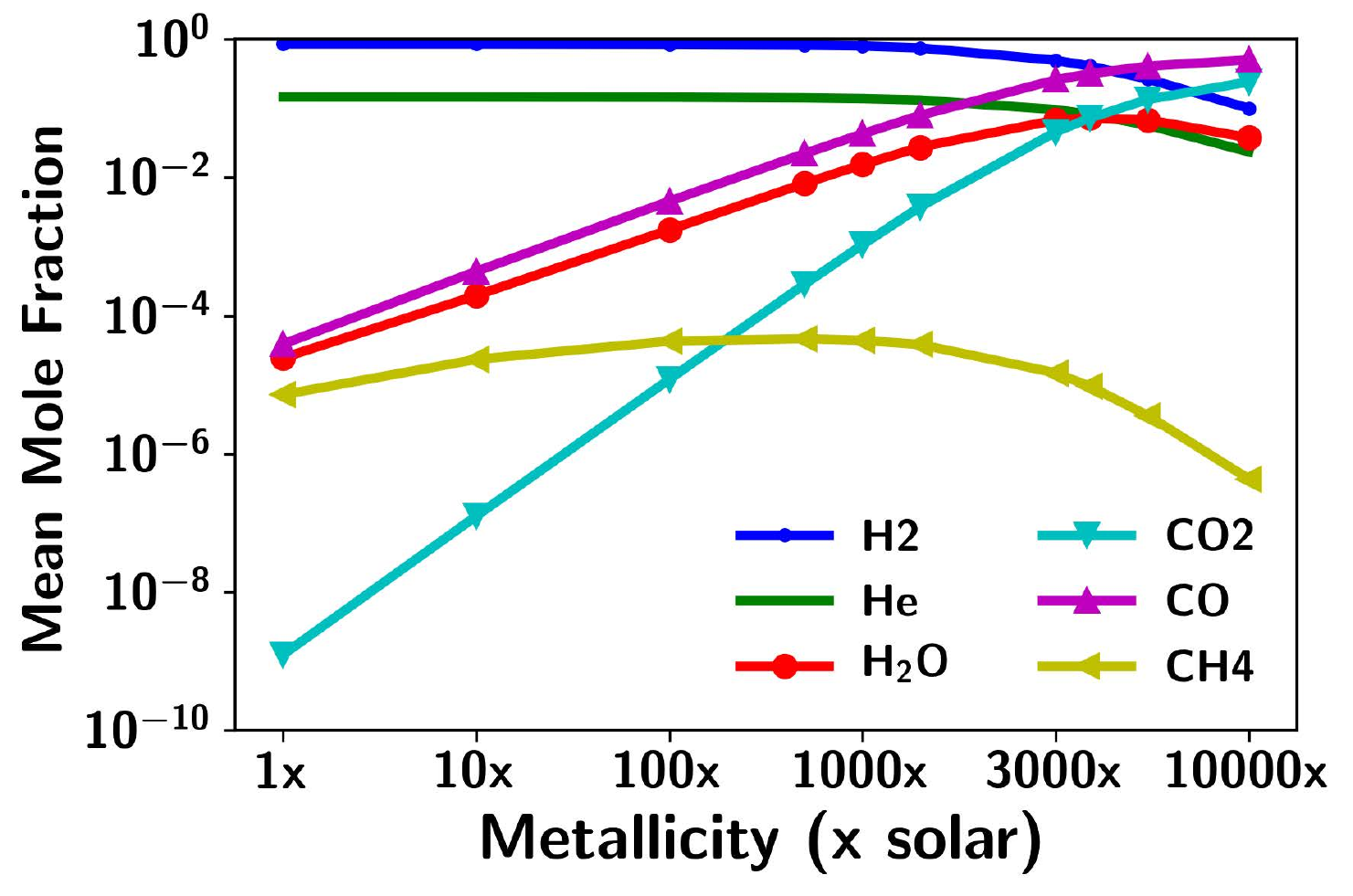}
\caption{\label{fig:mmw_vs_met_jmg}\small
Change in the mean mole fraction of spectrally-important chemical species, in the transmission spectra probed region (0.1 to 100 millibars) with change in planetary atmospheric metallicity. Model parameters for the calculations include: T = 1000 K, gravity = 10 m/s$\textsuperscript{2}$, C/O = 0.56, no clouds, no hazes, and all metallicities.}
\end{figure}

To address the question of distinguishing super-Earths from sub-Neptunes, we divided our spectra into two sets by MMW to coarsely represent the two types of exoplanets.  \cite{Benneke2013} found that an atmosphere with MMW $<$10 u will have a H/He fraction of at least 50\%, and they used 4 - 16 u to define the transition region between H$_2$- and H$_2$O-rich atmospheres.  We adopt a similar value of 10 u (with u denoting atomic mass units (AMU)), as illustrated in Figure \ref{fig:mmw_vs_met}, which shows MMW for our spectra as a function of their metallicities.  An analysis of the hydrogen abundance (H$_2$+H) as a function of MMW also revealed a visible break in the data centered on \sim 10 u (Figure \ref{fig:h_vs_mmw}). From Figure \ref{fig:mmw_vs_met_jmg}, which shows the mean molar fraction of spectrally significant atmospheric chemical species as a function of atmospheric metallicity, we can see that this finding is consistent with the increasing prominence of heavier chemical species with increasing metallicity. We therefore assigned spectra with MMW $\leq$10 u to the sub-Neptune category  and spectra with MMW $>$10 u to the super-Earth category. We acknowledge that \cite{Kite2020} showed that it is theoretically possible to produce sub-Neptunes with H\textsubscript{2}O primary atmospheres (MMW > 10 u), and that arguments have been made for H\textsubscript{2}-rich secondary atmospheres (MMW < 10 u) on rocky exoplanets \citep{Swain2021}; however, given the MMW distribution of our model spectra (Figure \ref{fig:mmw_vs_met}), the corresponding atmospheric hydrogen (H\textsubscript{2}+H) fractions (Figure \ref{fig:h_vs_mmw}), and the findings of \cite{Lopez2014} and \cite{Benneke2013}, we find the 10 u division to be appropriate for our model set.

Figure \ref{fig:h_vs_mmw} also shows a small cluster at 48.5 \% hydrogen abundance corresponding to MMW of 7.96 associated with T = 400 and 500 K, and C/O ratio of 0.35. Subsolar C/O ratio (0.35) provides more oxygen in the atmosphere, thus favoring formation of CO$_2$ at higher metallicities (>100x solar) and lower temperatures, thereby increasing the atmospheric MMW as shown by this cluster.

\section{Filter Definition} \label{sec:filters}
To ensure a thorough investigation of these spectra, we explored a wavelength range of 0.2 - 30 $\mu$m with R = 20, which results in 104 filters with 5253 filter pair combinations.  These wavelengths represent the range over which it is possible to achieve a high SNR using current instrumentation technologies. This range encompasses the wavelengths that are expected to be useful for transmission observations by operational telescopes like \emph{HST} as well as {\jwst} and other near-term telescope concepts (\textit{e.g.,} ARIEL \citep{ARIEL2020}, HabEx \citep{HabEx2020}, LUVOIR \citep{LUVOIR2019}, Origins Space Telescope \citep{Battersby2018}, and MIRECLE \citep{Staguhn2019}). Additionally, this filter set covers multiple absorption features which can serve as possible habitability and/or biosignature indicators (\textit{e.g.,} H$_2$O, CH$_4$, CO$_2$, N$_2$O, O$_2$, O$_3$).

\section{Color Metric and Analysis Methods} \label{sec:metric}

\subsection{Updates to \texttt{colorcolor} Analysis Framework} \label{sec:sub_color_updates}
We based our analysis framework and methods off of those described by \cite{Batalha2018}, whose effort developed the \hyperlink{https://github.com/natashabatalha/colorcolor}{\texttt{colorcolor}} code and analysis framework. They used this framework to apply supervised classification multivariate analysis methods to statistically evaluate 15 \emph{WFIRST}-like filter combinations against simulated spectra for 9120 extrasolar giant planets. This framework produces a database of color-color calculations for each filter combination and for each simulated spectrum ($n$ filters $\times$ $m$ spectra). This database can then be subjected to a simple correlation analysis and/or multivariate analysis to investigate the viability of color-color analyses for the classification of planets based on physical properties (\textit{e.g.}, atmospheric metallicity or temperature).  Moreover, the framework allows for the evaluation of several different types of classification algorithms, including Linear Discriminant Analysis (LDA), K-Nearest Neighbor (KNN), Classification and Regression Trees (CART), Multiclass Logistic Regression, a na\"\i ve Bayes classifier, and Support Vector Machines. The metric used to assess the success of an algorithm was the mean and standard deviation of the accuracy after a k-fold cross validation test \citep{Kohavi1995}. Of the six algorithms tested by \cite{Batalha2018}, only the LDA and CART yielded fruitful results for reflectance spectra of a population of extrasolar giants that included cloudy atmospheres evaluated against WFIRST filters. Given the high susceptibility of CART to overfitting, for this work we chose to limit this initial evaluation to the more straightforward LDA algorithm.

We have adapted the \texttt{colorcolor} framework to conduct a similar statistical investigation for transmission (rather than reflectance) spectra, with the goal of identifying correlations between transmission wave bands and atmospheric MMW for numerous filter combinations. Minor modifications to the framework included adding a script that parses the ATMO spectrum files into a Python Pandas database, as well as adding an interface to generate a set of filters from a specified minimum and maximum wavelength and resolution. We have also added functions to facilitate the evaluation and plotting of the Pearson correlation matrix and the within- and between-group variances of the labeled groups.

The primary change that we implemented to the \texttt{colorcolor} code was a modification of the color comparison metric from reflectance, which simply entails the flux ratios, to a calculation relevant for transmission.  To this end, we used Python's \texttt{scikit-learn} \texttt{LinearDiscriminantAnalysis} function to evaluate several different transmission color metrics.

\subsection{Filter Pair Down-selection} \label{sec:sub_downselect}
The LDA essentially represents an Eigen analysis that decreases data dimensionality and aims to maximize separation between labeled groups of dependent variables.  Since we are looking to estimate MMW using a minimal set of filter combinations, MMW is our dependent variable, while our independent variables are the transmission metric calculated for each filter pair.  The LDA produces weighting coefficients for each of the independent variables (in this case, our filter pair calculations), and these coefficients indicate the variables' contribution along the axis of separation.  These weighting coefficients can be used to prioritize filter pairs based on their expected utility in separating MMW groups.  The Python LDA function also returns a prediction of the group (in our case, MMW) of each spectrum based on the values of the filter pair transmission metrics.  The accuracy of this prediction served as our evaluation criterion for the different transmission metrics.

The primary change that we implemented to the \texttt{colorcolor} code was a modification of the color comparison metric from reflectance, which simply entails the flux ratios, to a calculation relevant for transmission.  To this end, we used Python's \texttt{scikit-learn} \texttt{LinearDiscriminantAnalysis} function to evaluate several different transmission color metrics.

Given our goal of identifying a small subset of filter pairs (ideally 3 or fewer pairs for instrument simplicity) that can facilitate distinguishing super-Earths from sub-Neptunes and the thousands of filter pair combinations available within our filter set, a method was required for identifying the most useful filter pairs based on our transmission metric.  As described previously, the LDA produces weighting coefficients that indicate the relative utility of the independent variables (\textit{i.e.,} filter pairs) in separating the dependent variable groups (\textit{i.e.,} MMW); the LDA function thereby provides one means of identifying useful filter pairs.  A correlation analysis produces the Pearson correlation coefficients between each pair of variables in the data set, and these coefficients provide another means of identifying which filter pairs are most correlated with MMW and should, therefore, be more useful for estimating atmospheric MMW.

\subsection{Filter Pair Evaluation process} \label{sec:sub_evaluation}
LDA is a supervised machine learning technique; it is ``supervised'' in that it accounts for the variable labels when maximizing separation between variable groups \citep{Batalha2018}.  The evaluation of machine learning algorithms involves the use of a training data set -- used by the machine learning algorithm to generate a model of the data -- and a test data set, which serves as ``new'' data against which to evaluate the models computed by the machine learning algorithm.  For this study, we used Python's \texttt{scikit-learn} \texttt{test\_train\_split} function to randomly split 75\% of our data into a training set and 25\% of our data into a test set against which the filter subsets were evaluated. 

In the interest of conducting a thorough analysis, we evaluated the performance of the LDA-favored filter pairs as well as the filter pairs most correlated with MMW.  Our filter pair evaluation process comprised the following steps.
\begin{enumerate}
	\item Split models into training data (75\%) and test data (25\%) subsets.
	\item Run the LDA for the training portion of a given data set (\textit{e.g.,} all models, cloud-free models, etc.) using all of the filter pair combinations and identify the three filter pairs with the highest LDA weighting coefficients.
	\item Compute the matrix of Pearson correlation coefficients for the training portion of the data set and identify the three filter pairs that were most correlated with MMW.
	\item Rerun the LDA for the training dataset using only the top three LDA-favored filter pairs and compute the LDA prediction accuracy for this three-filter-pair subset.
	\item Rerun the LDA for the training dataset using only the top three MMW-correlated filter pairs and compute the LDA prediction accuracy for this three-filter-pair subset.
	\item Use the LDA coefficients from both three-filter-pair training runs to predict the categories for each spectrum in the test data set and compute the prediction accuracies for the two different filter sets based on the MMW assignments for the test data set.
\end{enumerate}

\begin{figure*}[ht]
\includegraphics[width=1.0\linewidth]{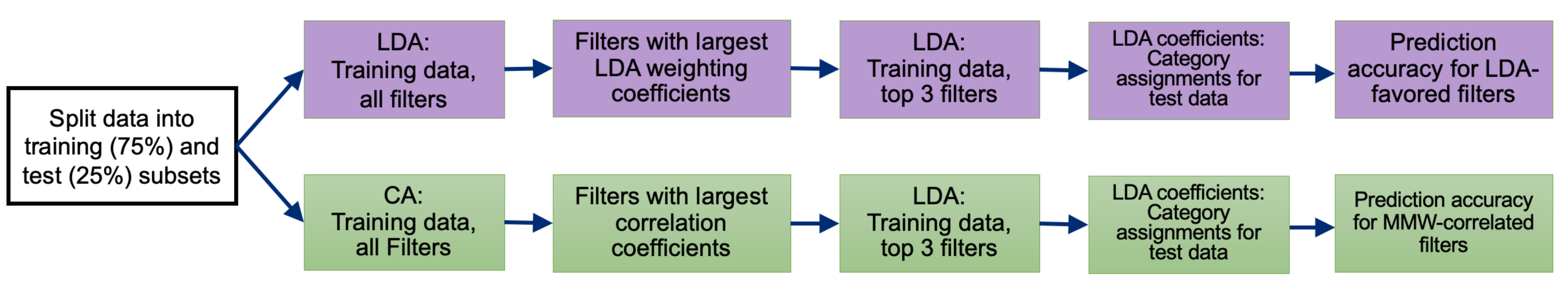}
\caption{\label{fig:lda_process}\small
Filter pair evaluation process.  A LDA and a correlation analysis (CA) are performed on the training data set using all of the filter pairs to determine the top three filter pairs for each down-selection method.  The training data are used to perform the LDA for the top three LDA-favored and MMW-correlated filter pairs, and the LDA coefficients are then used to predict each model spectrum's category based on these coefficients and the "color-color" calculation for the spectrum.  These predictions are then compared to the MMW assignments for the test spectra to calculate the prediction accuracy for the filter set.}
\end{figure*}

Figure \ref{fig:lda_process} illustrates this process flow.

\subsection{Transmission Color Metrics} \label{sec:sub_metrics}

In order to preserve the relative differences between the various spectra, we explored a simple transit depth subtraction metric (Equation \ref{eq:eq3}) as well as a transit depth ratio metric (Equation \ref{eq:eq4}).  
\begin{equation}\label{eq:eq3}
\Delta d = d\textsubscript{$2$} - d\textsubscript{$1$}
\end{equation}

\begin{equation}\label{eq:eq4}
\Delta d = \frac{d\textsubscript{$2$}}{d\textsubscript{$1$}}
\end{equation}

Equation \ref{eq:eq3} encapsulates the relationship of the absorption features to the radius of the host star, while Equation \ref{eq:eq4} strictly captures the comparative signal strength between the wave bands.  We discuss the performance for the different transmission metrics in the following section.

\section{Results \& Discussion} \label{sec:results}

\subsection{Analysis Results}

Our primary gauge as to the utility of a given transmission metric/filter pair combination is the prediction accuracy resulting from the second LDA run with the selected filter subset.  This prediction accuracy depends on the transmission metric, but also the method of filter selection (LDA vs correlated), and the number of filter pairs (with more filters providing improved accuracy).  Our central goal in this study was to determine whether there exists a reasonable number of filter pairs that provides a high prediction accuracy for differentiating super-Earths from sub-Neptunes based on atmospheric MMW.  

\subsubsection{Top Filter Pairs and Transmission Metric Performance}
\label{sec:results_metric_comp}

\begin{deluxetable}{c|c|c|c}
\tablecaption{Transmission Metric Performance Comparison for Super-Earths/Sub-Neptunes MMW Groups and All Model Parameters.  \label{tab:metricComp}}
\tablehead{\colhead{Transmission} & \colhead{Filter} & \colhead{Top Filter} & \colhead{Accuracy}\\ 
\colhead{Metric} & \colhead{Subset} & \colhead{Pairs ($\mu$m)} & \colhead{(\%)}
}
\startdata
 \multirow{6}{*}{$d_2 - d_1$} & & 0.205, 0.216 &\\ 
& LDA-favored & 0.216, 0.226 & 88.3\\
& & 0.216, 0.238 &\\\cline{2-4}
& & 1.187, 1.246 &\\
& Correlated & 3.308, 4.021 & 93.1\\
& & 1.130, 1.309 &\\
 \hline
 \multirow{6}{*}{$d_2/d_1$} & & 4.222, 24.452 &\\ 
& LDA-favored & 4.655, 26.958 & 87.8\\
& & 1.515, 2.468 &\\\cline{2-4}
& & 1.187, 1.246 &\\
& Correlated & 3.308, 4.021 & 93.2\\
& & 1.130, 1.309 &\\
\enddata
\begin{tablenotes}
    \small
    \item Note: ``Correlated'' refers to the MMW-correlated filter pairs.
\end{tablenotes}
\end{deluxetable}

\begin{figure}[t]
\includegraphics[width=1.0\linewidth]{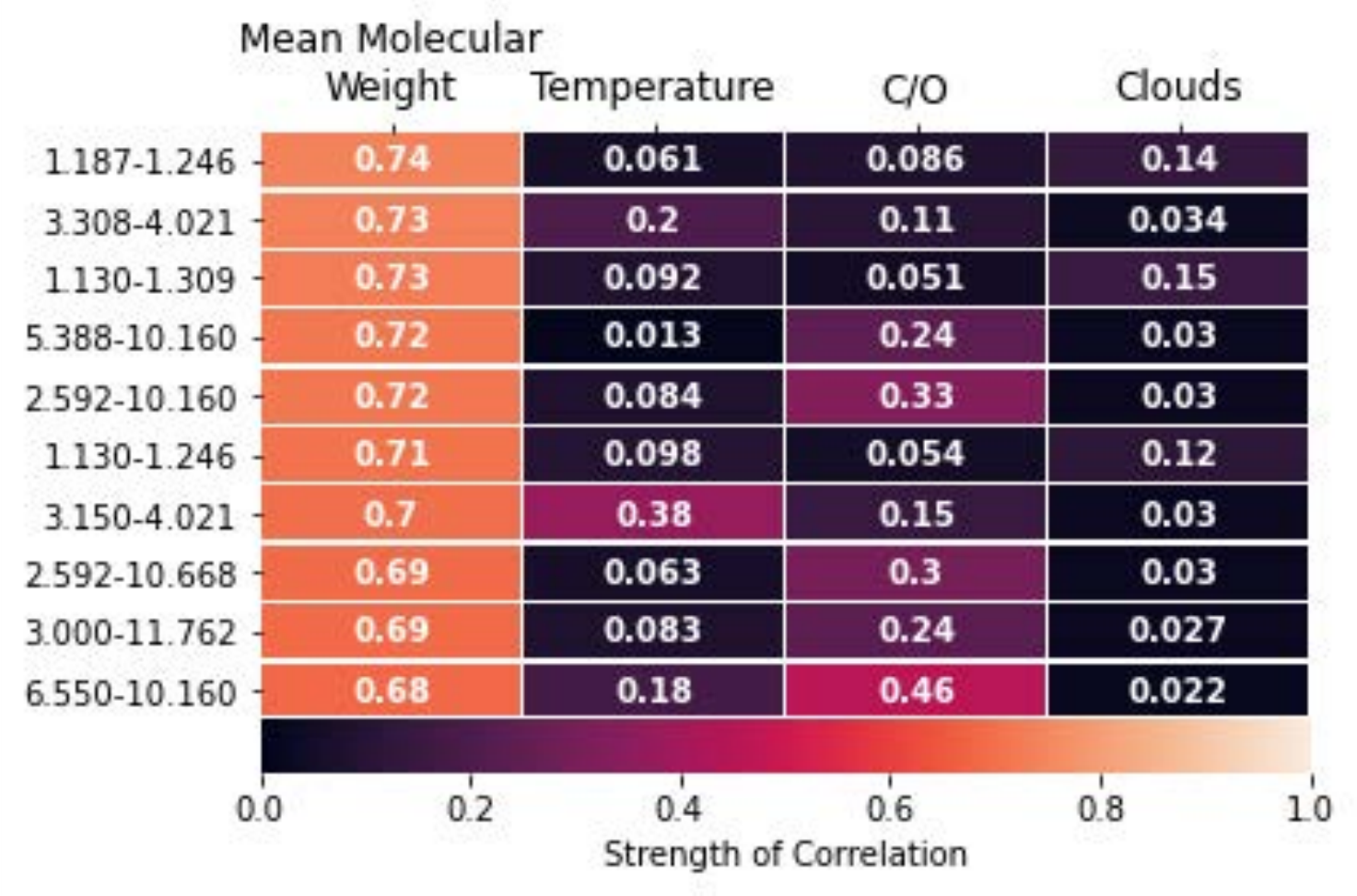}
\caption{\label{fig:corr_diff}\small
The top ten MMW-correlated filter pairs for the \emph{difference} metric, shown in comparison with the corresponding correlation coefficients for the temperature, C/O ratio, and cloud parameters.  The pairs that are most correlated with MMW are not strongly correlated with temperature, C/O ratio, or clouds, indicating that separate filter pairs could be used to estimate these parameters.  The top ten MMW-correlated filter pairs for the ratio metric comprise the same subset of filters, with comparable correlation coefficients, though the pair combinations and correlation order are slightly different.
}
\end{figure}

\begin{figure}[h]
\includegraphics[width=1.0\linewidth]{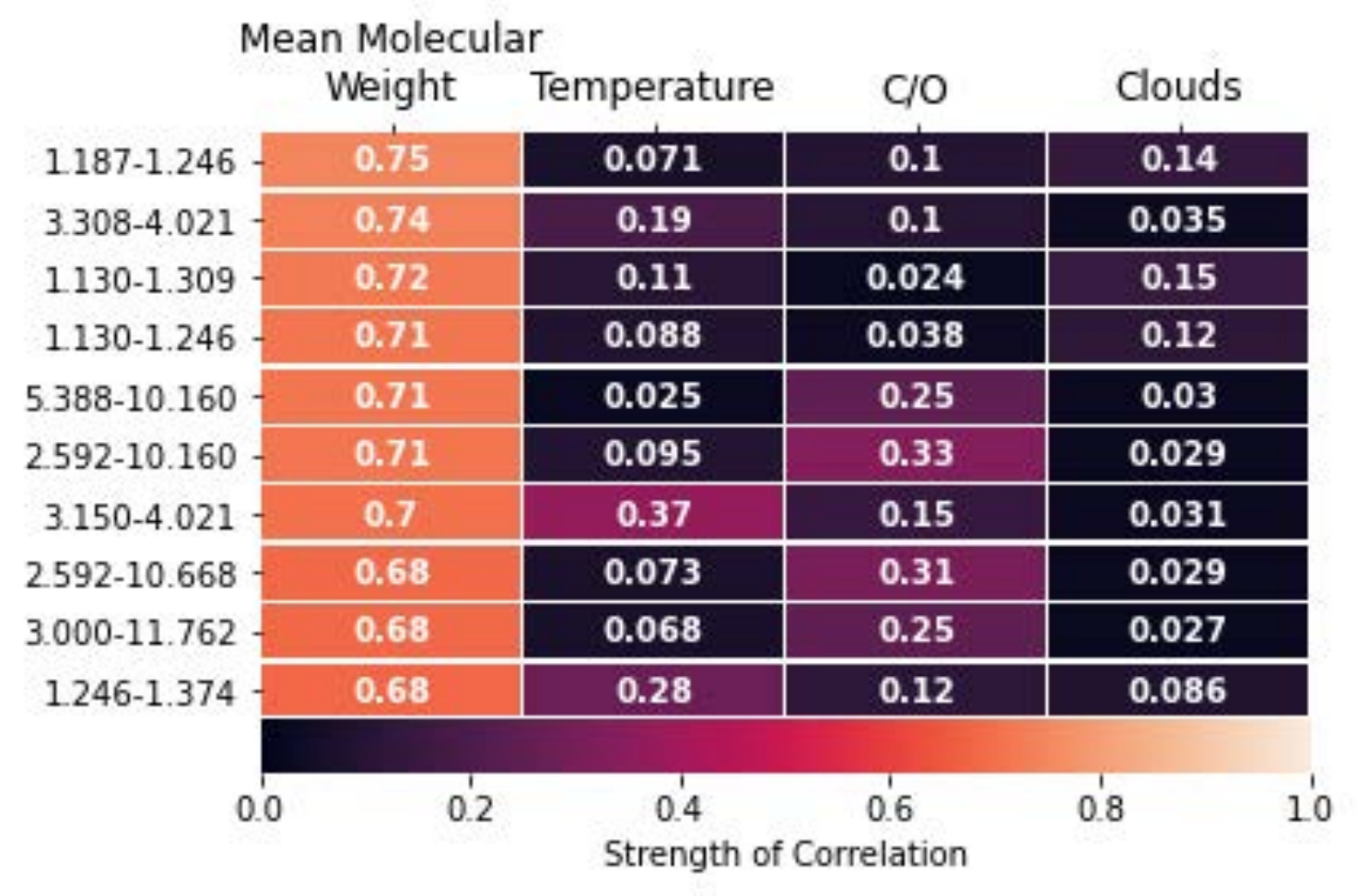}
\caption{\label{fig:corr_ratio}\small
The top ten MMW-correlated filter pairs for the \emph{ratio} metric, shown in comparison with the corresponding correlation coefficients for the temperature, C/O ratio, and cloud parameters.  The most-correlated filter pairs and correlation coefficients are similar to those for the difference metric, providing comparable performance in the LDA for the two metrics.}
\end{figure}

Table \ref{tab:metricComp} provides a comparison of the three transmission metrics described in the previous section for the top three LDA and MMW-correlated filter pairs for the two MMW groups described in Section \ref{sec:simspec}. The accuracy represents the prediction accuracy of each metric and filter pair subset for the full model set described in Section \ref{sec:simspec}. We see that the difference (Equation \ref{eq:eq3}) and ratio (Equation \ref{eq:eq4}) metrics provide comparable performance for three filter pairs for both the LDA-favored and MMW-correlated filter sets.  We also see that the MMW-correlated filters are the same for both metrics and provide better performance than the LDA-favored filters.  Figure \ref{fig:corr_diff} shows the top ten MMW-correlated filter pairs for the difference metric for the model set described in Section \ref{sec:simspec}. We can see that the pairs that are most correlated with MMW are not strongly correlated with temperature, C/O ratio, or clouds, indicating that separate filter pairs could be used to estimate these parameters. The top ten MMW-correlated filter pairs for the ratio metric comprise the same subset of filters as for the difference metric, and the MMW correlation strength is comparable, though the pair combinations and correlation order are slightly different, as can be seen in Figure \ref{fig:corr_ratio}.

\begin{figure}[t]
\includegraphics[width=1.0\linewidth]{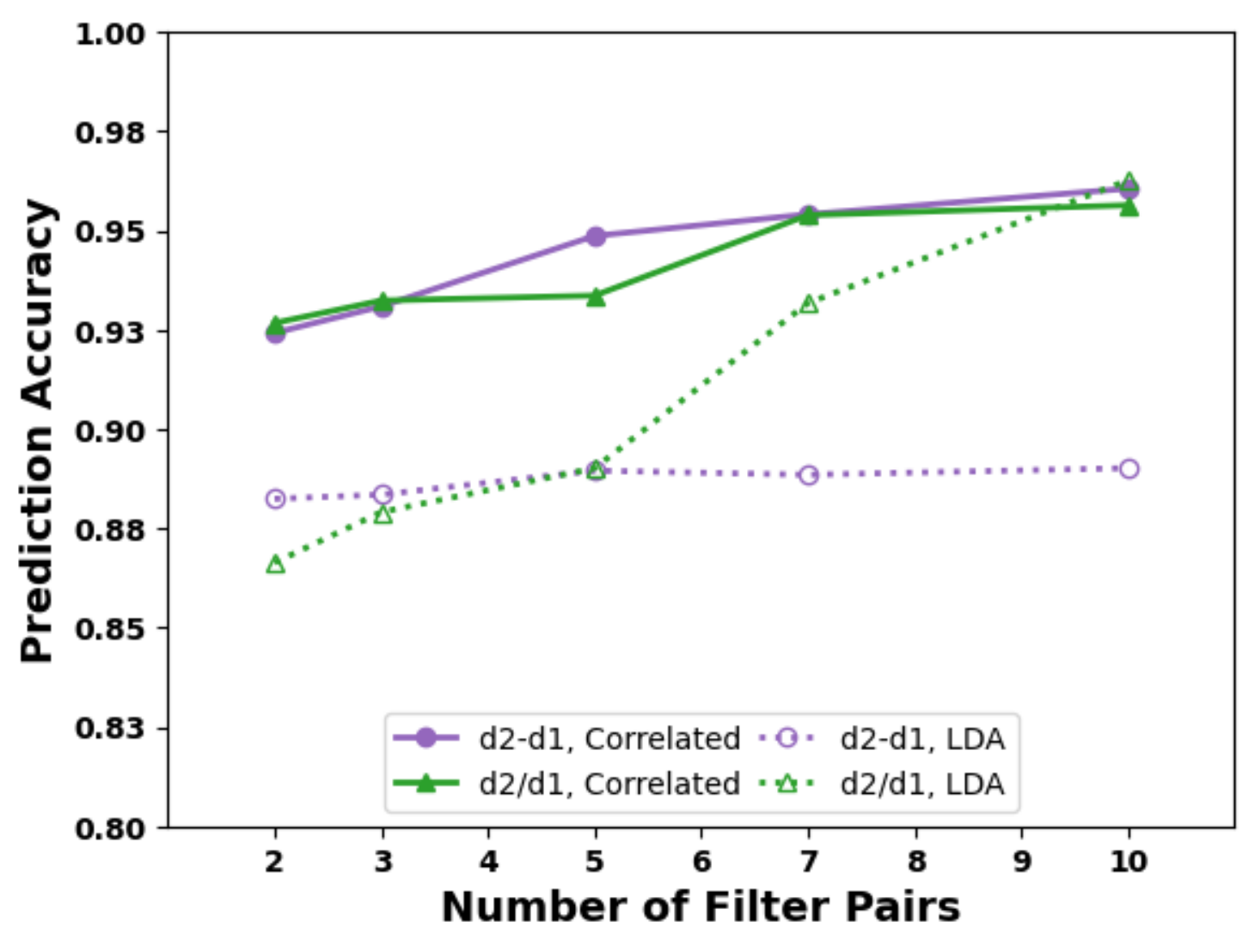}
\caption{\label{fig:acc_vs_num_filts}\small
Prediction accuracies for the three transmission metrics for different numbers of MMW-correlated and LDA-favored filter pairs. The difference and ratio metrics perform comparably, particularly for smaller numbers of filter pairs, and the correlated filters generally outperform the LDA-favored filters for up to ten pairs. Additionally, increasing the number of correlated filter pairs from two to ten does not significantly increase the prediction accuracy for the difference and ratio metrics and the correlated filters.
}
\end{figure}

Given the comparable performance of the difference and ratio metrics shown in Table \ref{tab:metricComp} when using three filter pairs, we also evaluated the metrics against different numbers of filter pairs. Figure \ref{fig:acc_vs_num_filts} shows the prediction accuracy for all three metrics as a function of the number of filter pairs.  

We limited the number of filter pairs to ten for this analysis because more than ten pairs was assumed to be impractical for observational use.  From this figure, we can see that for the two MMW groups described in Section \ref{sec:simspec}, the difference and ratio metrics demonstrate similar performance, particularly for smaller numbers of filter pairs, and that \textbf{as few as two filter pairs produces a prediction accuracy of $\sim$93\% for noise-free spectra of isothermal atmospheres in chemical equilibrium.  We also see that} for up to ten filter pairs, the MMW-correlated filters perform comparably to or better than the LDA-favored filters.  Additionally, for the difference and ratio metrics, increasing the number of filter pairs to ten improves the prediction accuracy by only a few percent. Moreover, while the difference and ratio metrics perform comparably overall, the difference metric performs slightly better than the ratio metric in most circumstances.

The consistent performance of the difference and ratio metrics with number of filters is due to the similar correlation coefficients of the top ten filter pairs, which indicates that the information content for these filter pairs is comparable (see Figure \ref{fig:corr_diff}).  Likewise, the performance difference between the difference and ratio metrics is due to the filter pairs that are most strongly correlated with MMW, which are similar for the two metrics, with slight differences in the correlation coefficients.  The variation in MMW prediction accuracy for the two metrics comes from the different filter pairs that are used in the LDA based on the correlation strength.  Correspondingly, the performance for ten filter pairs is nearly identical for the two metrics because the set of filter pairs is the same, excepting the tenth filter pair, thereby providing essentially identical information to the LDA.  Furthermore, the top ten filter pairs for both metrics are primarily triggering on a couple of spectral features, like the 1.2-$\mu$m water feature and the 3.3-$\mu$m methane feature, again limiting the information added by these supplemental filters.

Overall, the performance of the difference and ratio metrics are comparable, and there is a strong overlap among the most MMW-correlated, and highest-performing, filter pairs. However, based on Figure \ref{fig:acc_vs_num_filts} and Table \ref{tab:metricClouds}, for the purpose of differentiating super-Earths from sub-Neptunes, we recommend the transit depth difference metric (Equation \ref{eq:eq3}) in conjunction with a combination of the filter pairs listed in Figure \ref{fig:corr_diff}. Table \ref{tab:filters} lists the wavelength limits for the aforementioned filters.

While our population of model spectra includes several simplifying assumptions (e.g., isothermal and chemical equilibrium), as shown in Figures \ref{fig:mmw_vs_met} and \ref{fig:h_vs_mmw}, our model atmospheres span a range of MMWs, particularly in the super-Earth category.  We therefore expect this method to be able to distinguish between sub-Neptunes and super-Earths with $\sim$90\% accuracy for a variety of secondary atmospheres, such as N$_2$-, O$_2$-, and CO$_2$-dominated atmospheres.

\begin{deluxetable}{c|c|c}
\tablecaption{Wavelength Limits for MMW-Correlated Filters \label{tab:filters}}
\tablehead{\colhead{Filter Center} & \colhead{Minimum} & \colhead{Maximum} \\
\colhead{($\mu$m)} &\colhead{($\mu$m)} & \colhead{($\mu$m)}
}
\startdata
1.187 & 1.158 & 1.216\\ 
1.246 & 1.216 & 1.277\\
3.308 & 3.227 & 3.389\\
4.021 & 3.923 & 4.119\\
1.130 & 1.103 & 1.158\\
1.309 & 1.277 & 1.341\\
5.388 & 5.257 & 5.520\\ 
1.0160 & 9.912 & 10.408\\
2.592 & 2.529 & 2.655\\
3.150 & 3.073 & 3.227\\
1.0668 & 10.408 & 10.928\\
3.000 & 2.927 & 3.073\\
1.1762 & 11.475 & 12.048\\
6.550 & 6.309 & 6.709\\
\enddata
\end{deluxetable}

\subsubsection{Trends in Misclassification of Spectra}
\label{sec:results_misclass}

\begin{figure}[t]
\includegraphics[width=1.0\linewidth]{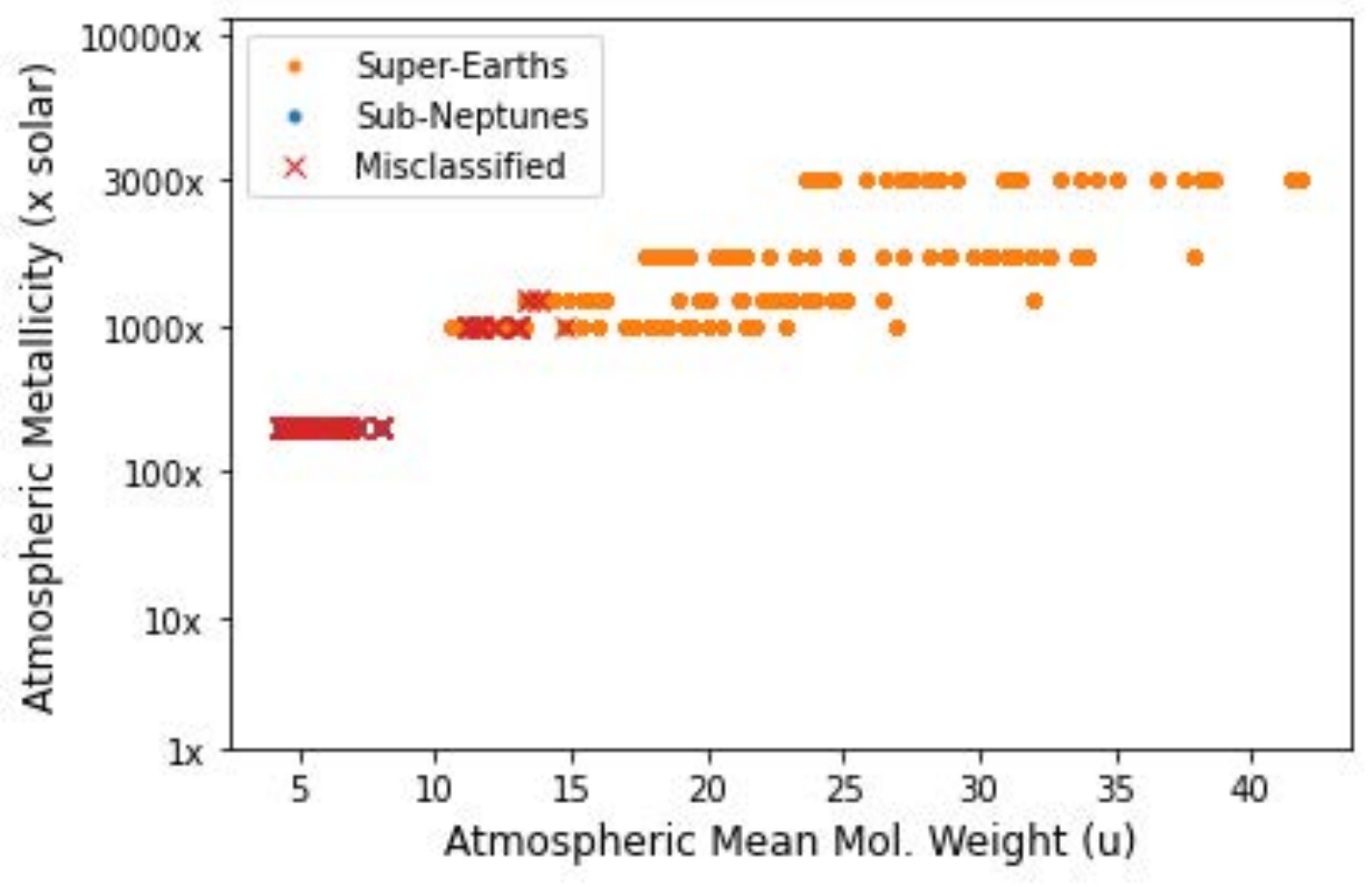}
\caption{\label{fig:misclass_met_mu}\small
Here we show Figure \ref{fig:mmw_vs_met} with misclassified spectra illustrated via red ``x''s.  It can be seen that all of the misclassified spectra are of MMW<15 u, and that misclassifications fall heavily in the sub-Neptune category.  This is due to the smaller percentage of sub-Neptune models (20.1\%) included in our data set compared to the percentage of super-Earth models (Table \ref{tab:confusionAll}).
}
\end{figure}

\begin{deluxetable}{c|c|c||c}
\tablecaption{MMW Prediction Confusion Matrix for All Models (Difference Metric, MMW-Correlated Filters). \label{tab:confusionAll}}
\tablehead{\colhead{} & \colhead{Predicted} & \colhead{Predicted} & \colhead{Total}\\
\colhead{} & \colhead{Super-Earth} & \colhead{Sub-Neptune} & \colhead{Actual}
}
\startdata
Actual & \cellcolor[HTML]{7be97d} & \cellcolor[HTML]{F39989} & \multirow{2}{*}{11472}\\
Super-Earth & \multirow{-2}{*}{\cellcolor[HTML]{7be97d}\textbf{11317}} & \multirow{-2}{*}{\cellcolor[HTML]{F39989}\emph{155}} & \\
\hline
Actual & \cellcolor[HTML]{F39989} & \cellcolor[HTML]{7be97d} & \multirow{2}{*}{2892}\\
Sub-Neptune & \multirow{-2}{*}{\cellcolor[HTML]{F39989}\emph{919}} &  \multirow{-2}{*}{\cellcolor[HTML]{7be97d}\textbf{1973}} & \\
\hline
\hline
Total & \multirow{2}{*}{12236} & \multirow{2}{*}{2128} & \\
Predicted &  &  & \\
\enddata
\begin{tablenotes}
    \small
    \item The matrix diagonal (green, bold font) shows the numbers of correctly-classified planets for each category, while the off-diagonal (red, italic font) shows the number of incorrectly-classified planets for each category.
\end{tablenotes}
\end{deluxetable}

\begin{figure*}[t]
\includegraphics[width=1.0\linewidth]{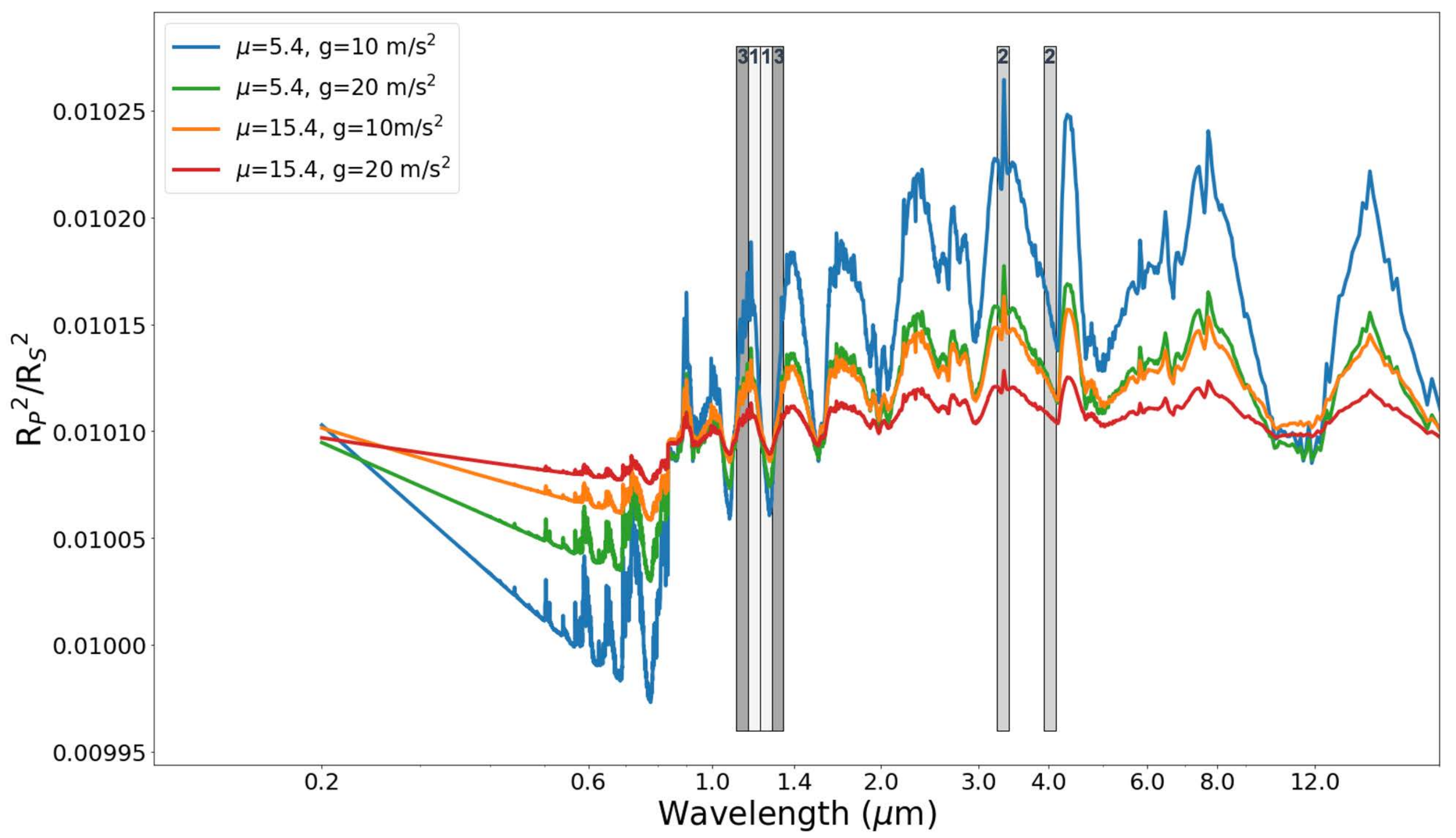}
\caption{\label{fig:spec_with_filters}\small
Four model spectra are shown with the top three MMW-correlated filter pairs to demonstrate the degeneracy between MMW and gravity near the super-Earth/sub-Neptune transition region between $\sim$10-15 u. It can be seen that the low-MMW, high-gravity spectrum looks very similar in the near-infrared wavelengths to the higher-MMW, low-gravity spectrum. This is to be expected, since the scale height is inversely proportional to both of these parameters, ergo they are coupled in their impacts on spectral features.  We can see that the top three filter pairs are triggering on known spectral features -- the first and third pairs trigger on the water feature centered on 1.1 $\mu$m, and the second pair triggers on the 3.3-$\mu$m methane feature.  Additional model parameters for all four spectra: T = 600 K, C/O = 0.70, cloud = 0 cm$^2$/g, haze = 1x Rayleigh scattering.}
\end{figure*}

In addition to the overall performance of the metrics for a given filter set, it is important to understand which spectra were classified incorrectly by the model and whether there are trends in the misclassifications.  Figure \ref{fig:misclass_met_mu} shows Figure \ref{fig:mmw_vs_met} with red ``x''s indicating spectra that were misclassified for the MMW-correlated filter set (top three).  As shown in Figure \ref{fig:misclass_met_mu}, all of the higher-MMW spectra were classified correctly, while the super-Earth/sub-Neptune misclassifications were confined to MMW<15 u, and primarily to the sub-Neptune category (MMW<10 u). This is additionally demonstrated by the confusion matrix presented in Table \ref{tab:confusionAll}, where we can see that a higher percentage, 31.8\%, of sub-Neptunes were misclassified, compared to 1.4\% of super-Earths. For MMW<15 u, the LDA achieved a prediction accuracy of 81.4\%, and for MMW<10 u (sub-Neptunes), it achieved a prediction accuracy of 68.2\%. The decreased MMW-prediction performance for sub-Neptunes is due to the smaller percentage (20.1\%) of sub-Neptune models in our data set relative to the percentage of super-Earth models (Table \ref{tab:confusionAll}), resulting in a smaller representative sub-Neptune population on which the LDA can train and compute feature separability.

\begin{figure}[h]
\includegraphics[width=1.0\linewidth]{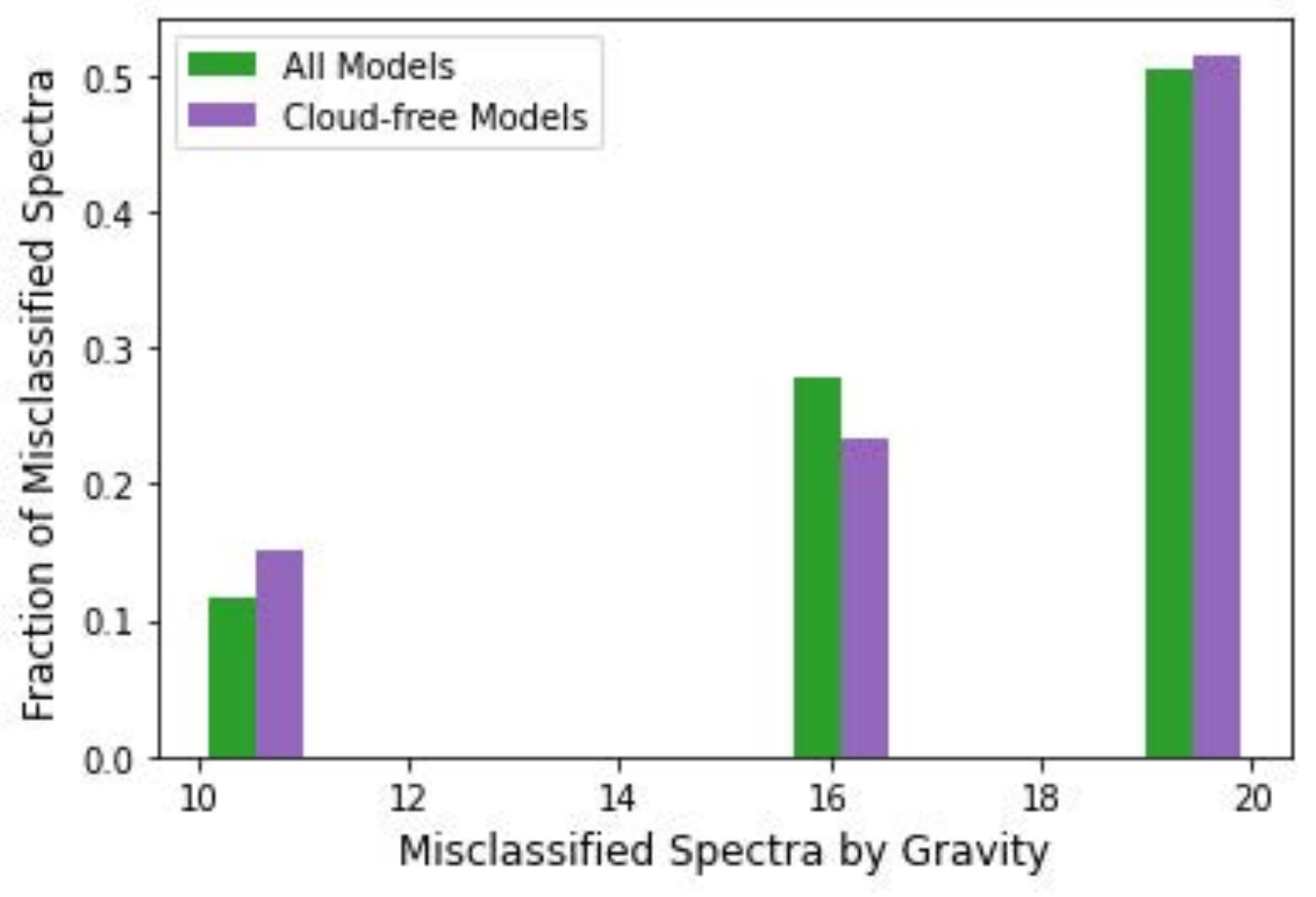}
\caption{\label{fig:misclass_grav}\small
Similar MMW-prediction performance can be achieved for both the full and cloud-free data sets.  We identify a trend in the impact of gravity on MMW-prediction accuracy, where we see increasing percentages of misclassified spectra with increasing gravity.  This is due to the muting effects of gravity on spectral features (\emph{e.g.,} \cite{Goyal2019}) that mimics a higher MMW atmosphere.
}
\end{figure}

Examining the relationship between MMW-prediction accuracy for the various model parameters, the strongest trend that we identified was that of the effect of gravity on MMW-prediction performance, which is due to the muting effects of gravity on spectral features (\emph{e.g.,} \cite{Goyal2019}). We can see from Figure \ref{fig:spec_with_filters} that for lower-MMW atmospheres (<$\sim$15 u), low-MMW/high-gravity spectra appear similar in the near-infrared wavelengths to higher-MMW/low-gravity spectra when other model parameters are the same.  This is to be expected, since the atmospheric scale height is inversely proportional to the MMW and the gravitational constant, and the impact of these two parameters on spectral features is therefore coupled. Figure \ref{fig:spec_with_filters} demonstrates this effect by showing four spectra with different MMWs and gravities, and we can see that the low-MMW/high-gravity spectrum looks very similar overall to the high-MMW/low-gravity spectrum, particularly in the near-infrared wavelengths encompassed by the top filter pairs.  In Figure \ref{fig:misclass_grav}, we see that increasing gravity negatively impacts MMW prediction for both the full and cloud-free model sets, with comparable performance impact between the two data sets.  We can also see in Figure \ref{fig:spec_with_filters} that the filters are triggering on known water (1.1 $\mu$m) and methane (3.3 $\mu$m) features in the near-infrared, and in particular the relative amplitudes and slopes of these features, which is consistent with the findings of \cite{Benneke2012}.  A weak correlation was found between misclassified spectra and temperature, indicating higher misclassification rates for cooler temperatures, though this was a secondary effect compared to that of the gravity parameter.

\subsubsection{Findings for Cloud-free Spectra}

Clouds are present on every solar system world that hosts a substantial atmosphere, and they have been detected for a number of exoplanets. Clouds are complex, with a variety of properties, and they provide broadband absorption and scattering, thereby complicating the interpretation of exoplanet spectra \citep{Line2016a,Barstow2020}. These effects are aggravated for transiting planets and transmission spectra, as the optical depth of condensates and hazes are significantly increased at the slant viewing geometry of transmission spectra, further complicating molecular abundance constraint calculations \citep{Fortney2005}. Most relevant to this study, clouds have muting effects on spectral features, caused by the reduction of the amplitude of spectral features due to the atmosphere becoming opaque below the cloud top, an effect that can be difficult to distinguish from a high-MMW, low-scale height atmosphere \citep{Benneke2012,Benneke2013,Barstow2020}. 

Given the degenerate muting effects of clouds and MMW on spectral feature amplitude and the \cite{Batalha2018} findings on the negative impact of clouds on the success of reflectance color analysis, we also evaluated the transmission color analysis method against the cloud-free model set (cloud parameter = 0.0).  Table \ref{tab:metricClouds} provides a comparison of the cloud-free vs full model set prediction accuracies for the three metrics and the three most MMW-correlated filter pairs.  From this table, we can see that, unlike for reflectance photometry \citep{Batalha2018, Grenfell2020}, clouds have very little impact on the predictive performance of our transmission color analysis -- the super-Earth \slash sub-Neptune prediction accuracies are essentially the same between the two data sets. The very slight decrease in prediction performance that we see for some of the cloud-free cases is due to the number of training models being decreased by a factor of four when using only the cloud-free models, as is evident in a comparison of the confusion matrices in Tables \ref{tab:confusionAll} and \ref{tab:confusionNoCloud}.

\begin{deluxetable}{c|c|c|c}
\tablecaption{Transmission Performance for MMW-Correlated Filters \& Super-Earth/Sub-Neptune MMW Groups for All vs Cloud-free Models. \label{tab:metricClouds}}
\tablehead{\colhead{} & \colhead{Model Set} & \colhead{Top Filter} & \colhead{Accuracy} \\
\colhead{} &\colhead{} & \colhead{Pairs ($\mu$m)} & \colhead{(\%)}
}
\startdata
 \multirow{6}{*}{$d_2 - d_1$} & & 1.187, 1.246 &\\ 
& All Models & 3.308, 4.021 & 93.1\\
& & 1.130, 1.309 &\\\cline{2-4}
& & 1.187, 1.246 &\\
& Cloud-free & 1.130, 1.309 & 92.5\\
& & 3.308, 4.021 &\\
 \hline
 \multirow{6}{*}{$d_2/d_1$} & & 1.187, 1.246 &\\ 
& All Models & 3.308, 4.021 & 93.2\\
& & 1.130, 1.309 &\\\cline{2-4}
& & 1.187, 1.246 &\\
& Cloud-free & 3.308, 4.021 & 92.6\\
& & 1.130, 1.309 &\\
\enddata
\end{deluxetable}

\begin{deluxetable}{c|c|c||c}
\tablecaption{MMW Prediction Confusion Matrix for Cloud-free Models (Difference Metric, MMW-Correlated Filters). \label{tab:confusionNoCloud}}
\tablehead{\colhead{} & \colhead{Predicted} & \colhead{Predicted} & \colhead{Total}\\
\colhead{} & \colhead{Super-Earth} & \colhead{Sub-Neptune} & \colhead{Actual}
}
\startdata
Actual & \cellcolor[HTML]{7be97d} & \cellcolor[HTML]{F39989} & \multirow{2}{*}{3840}\\
Super-Earth & \multirow{-2}{*}{\cellcolor[HTML]{7be97d}\textbf{3810}} & \multirow{-2}{*}{\cellcolor[HTML]{F39989}\emph{30}} & \\
\hline
Actual & \cellcolor[HTML]{F39989} & \cellcolor[HTML]{7be97d} & \multirow{2}{*}{948}\\
Sub-Neptune & \multirow{-2}{*}{\cellcolor[HTML]{F39989}\emph{301}} & \multirow{-2}{*}{\cellcolor[HTML]{7be97d}\textbf{647}} & \\
\hline
\hline
Total & \multirow{2}{*}{4111} & \multirow{2}{*}{677} & \\
Predicted &  &  & \\
\enddata
\begin{tablenotes}
    \small
    \item The matrix diagonal (green, bold font) shows the numbers of correctly-classified planets for each category, while the off-diagonal (red, italic font) shows the number of incorrectly-classified planets for each category.
\end{tablenotes}
\end{deluxetable}

\begin{figure*}[ht]
\includegraphics[width=1.0\linewidth]{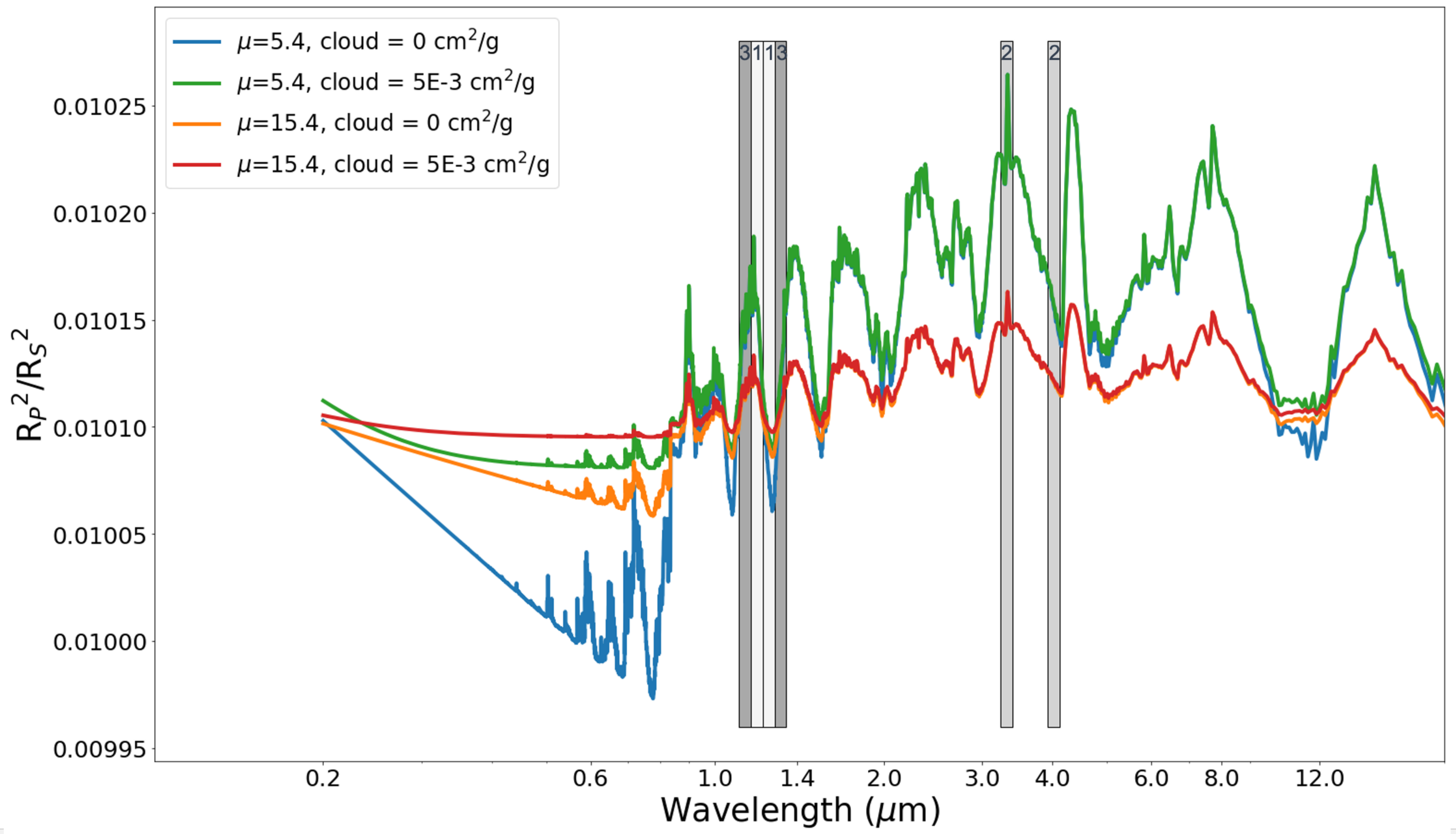}
\caption{\label{fig:mmw_clouds}\small
Four model spectra are shown with the top three MMW-correlated filter pairs to demonstrate the spectral effects of clouds for our models. It can be seen that the clouds obscure the depths of the spectral features, but that the muting of the features for the top three filters is driven more by the MMW.  Therefore, the relative differences between the peaks and troughs of the features for the different MMWs are \emph{comparatively} preserved between a cloud-free atmosphere and one with clouds for a given MMW.  Additional model parameters for all four spectra: T = 600 K, C/O = 0.70, g = 10 ms\textsuperscript{-2}, haze = 1x Rayleigh scattering.
}
\end{figure*}

The performance is consistent because the separability of the data into just two groups is similar for both the cloud-free and full data sets.  The impact of clouds is wavelength dependent, whereas the impact of mean molecular weight is not. Gray clouds impact the relative feature size of the near-infrared filters (1-1.4 $\mu$m) but not the mid-infrared features (3-4 $\mu$m), where clouds are transparent. Mean molecular weight impacts both.  Figure \ref{fig:mmw_clouds} shows a comparison of four model spectra with two different MMWs for both cloud-free (cloud particle cross section of 0 cm\textsuperscript{2}/g) and cloudy (cloud particle cross section of 5E-3 cm\textsuperscript{2}/g) atmospheres.  It can be seen in this figure that the first and third pairs (in the near-infrared) are affected by clouds, while the second pair is not, so the second pair is allowing for separability despite the presence of clouds.  Note also the very low correlation coefficients for the second filter pair with clouds in Figures \ref{fig:corr_diff} and \ref{fig:corr_ratio}.  

\begin{figure}[t]
\includegraphics[width=1.0\linewidth]{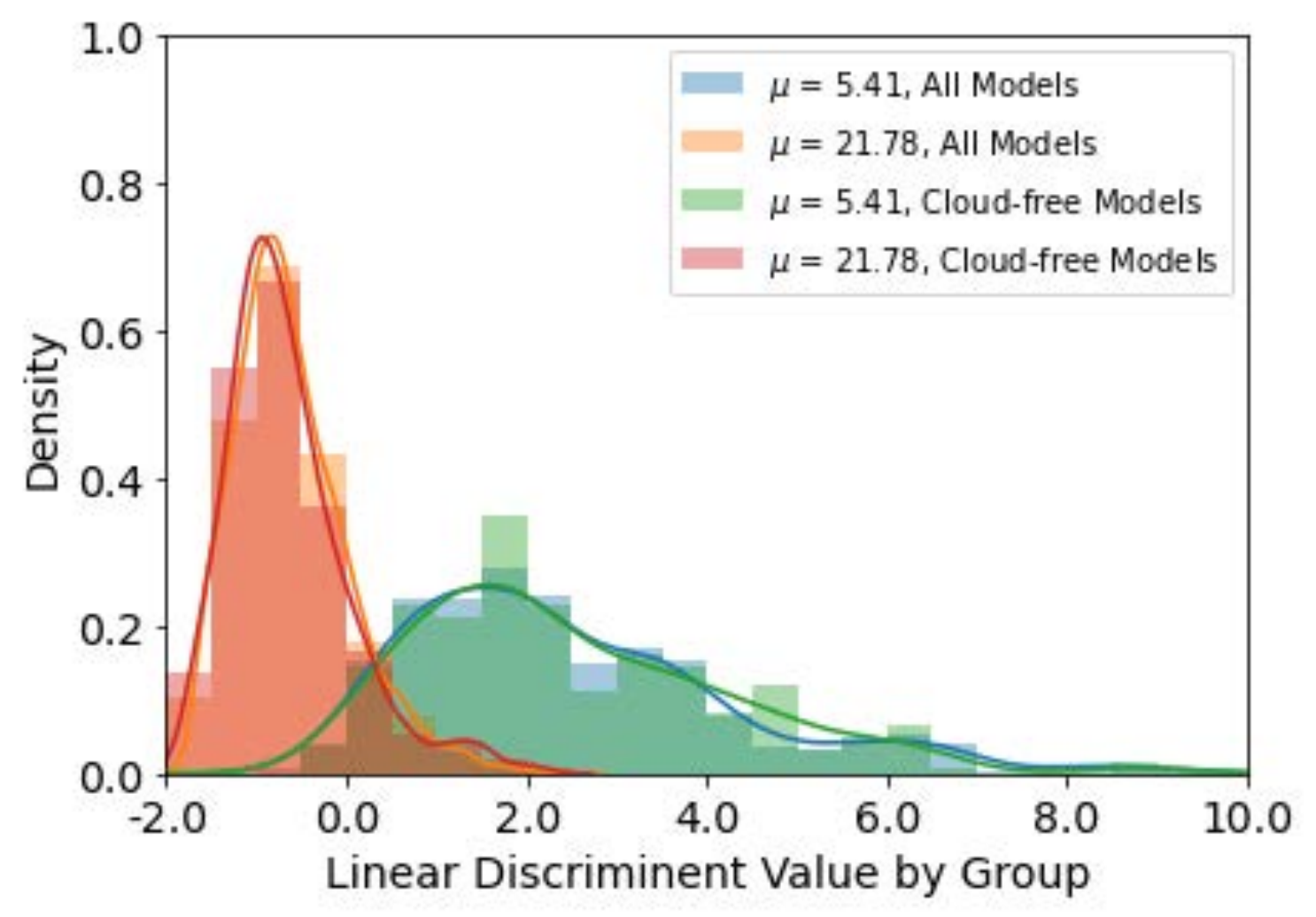}
\caption{\label{fig:lda_dual}\small
Linear discriminant values by MMW group for all models and for cloud-free models.  Note that there is strong overlap between the discriminant values for the two data sets, indicating that similar separability can be achieved by the LDA for both sets of data.  This results in the comparable performance of the transmission color analysis method for cloud-free data and for data that include cloudy spectra.
}
\end{figure}

\begin{figure*}[ht]
\includegraphics[width=1.0\linewidth]{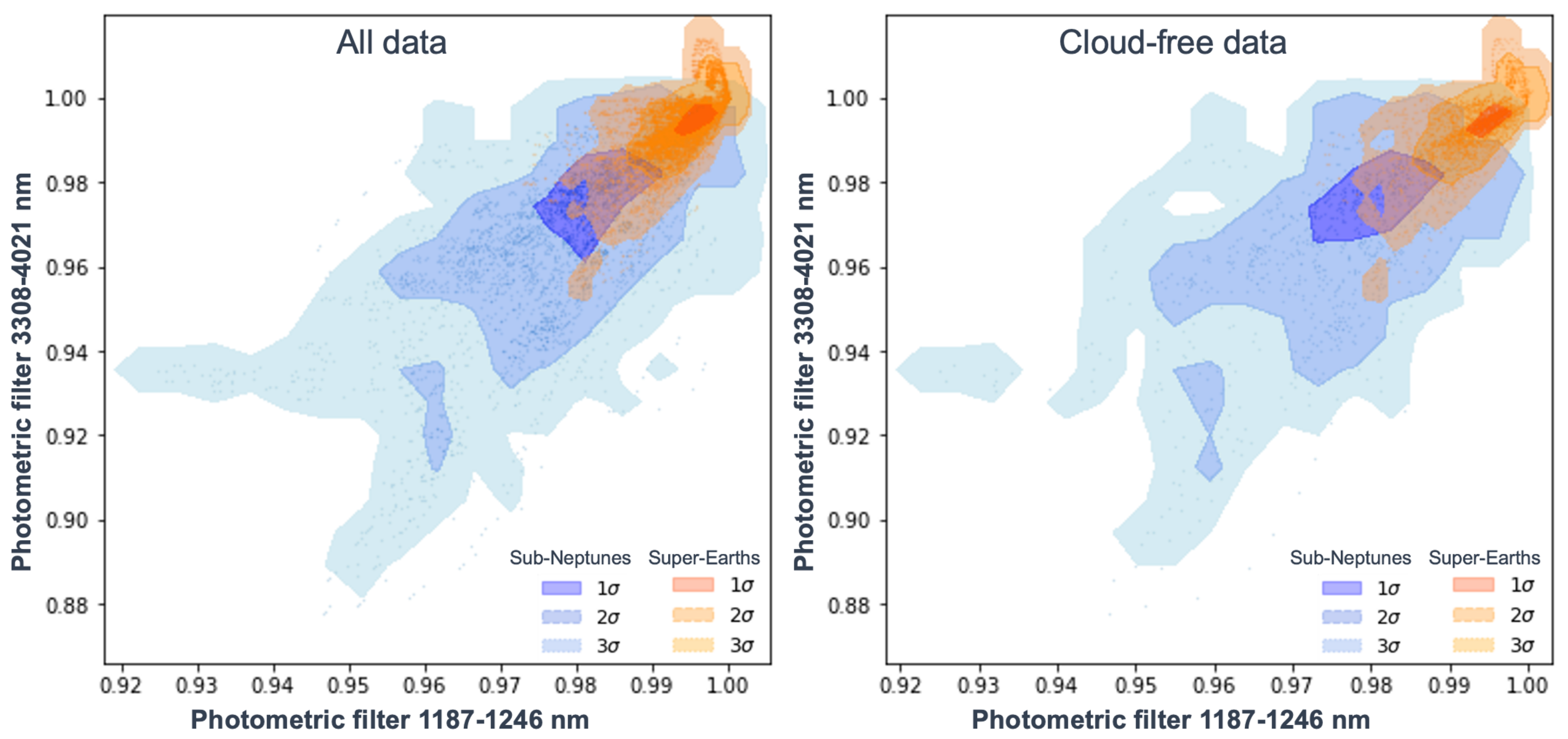}
\caption{\label{fig:color-color}\small
"Color" calculations and 1-, 2-, and 3-$\sigma$ contours for the difference metric for the full model set (left) and the cloud-free model set (right).  The distributions of the metric calculations and the standard deviation contours illustrate the tighter grouping of the super-Earth population and the overlap with the 2- and 3-$\sigma$ contours of the sub-Neptune population.  This results in the LDA favoring super-Earth assignments for planets that fall in the regions of overlap.  We also see that the distributions for the cloud-free metric calculations are extremely similar to those for the full model set, resulting in comparable performance between the cloud-free spectra and the full set of spectra.
}
\end{figure*}

Furthermore, Figure \ref{fig:lda_dual} depicts the linear discriminant values for the cloud-free and full data sets, and we can see that there is strong overlap between the linear discriminants for the two data sets as well as a consistently sized transition region between the two MMW groups. Based on the similar distribution of linear discriminant values, we would expect the MMW prediction accuracy to be comparable between the two data sets. This is also reflected in the numbers of misclassified spectra for the cloud-free data, shown in the confusion matrix in Table \ref{tab:confusionNoCloud}, where we see similar misclassification fractions that we saw for the full data set in Table \ref{tab:confusionAll}, as well as in Figure \ref{fig:color-color}, where we can see the distributions of the difference metric calculations for the two categories with their 1-, 2-, and 3-$\sigma$ contours.  These distributions show the tighter grouping of the super-Earth "color" calculations and the overlap with the 2- and 3-$\sigma$ portions of the sub-Neptune points.  Since the super-Earth population has a tighter grouping that overlaps with the higher standard deviation portions of the sub-Neptune population, it is more likely for a sub-Neptune in the overlap region to be classified as a super-Earth, which is consistent with our misclassification statistics.  Furthermore, we can see that the metric calculation distributions and standard deviation contours are extremely similar between the cloud-free and full model sets, which elucidates the comparable performance between these data sets.

We predict that this separability will be more strongly affected by clouds when we include more MMW groups and instrument noise. The former will likely yield increasing overlap in the linear discriminants than we see for just two groups, while the latter will hinder the ability of this technique to discriminate between the different chemistry regimes of our data set, which are more evident in our noise-free spectra; this may shift the best-performing filters into the optical wavelengths, as the Rayleigh scattering signature can help to constrain MMW for spectra of finite precision \citep{Benneke2012}.

\subsection{Implications for Future Observations}

With the high prediction accuracies for sub-Neptune and super-Earth MMW categorization, the color analysis methodology presented herein shows promise for use in future observations and characterization studies of these classes of exoplanets.  This is especially true considering that, in contrast to previous findings for reflected light photometry \citep{Batalha2018,Grenfell2020}, clouds do not impact our ability to determine atmopsheric MMW for noise-free spectra.  

The top ten filter pairs are covered by the {\jwst} instrument modes, and we have evaluated the MMW prediction performance for the relevant {\jwst} observing modes as well as for the \emph{Hubble Space Telescope}'s (\emph{HST}) Wide Field Camera 3(\emph{WFC3}), Table \ref{tab:jwstComp}, with all analyzed {\jwst} instrument modes and WFC3 resulting in >90\% prediction accuracy.  While the effects of instrument noise still need to be addressed, accounting for SNR considerations, we expect NIRISS/SOSS to be the recommended {\jwst} instrument mode for differentiating super-Earth and sub-Neptune atmospheres, which is consistent with the single-mode findings of \cite{Batalha2017-jwst_ic}.

The results discussed herein indicate that transmission "color" analysis has the potential to be useful for identifying with high confidence super-Earth and sub-Neptune atmospheres.  Moreover, the most useful filters fall in the near- and mid-infrared, a wavelength range that is both well understood and practical for current instrumentation technologies. This bodes well for validating this method using data from existing and near-term telescopes as well as for synthesizing this method with near-term observations from telescopes like {\jwst} and \emph{HST}, allowing for reconnaissance observations that can guide the observation planning (instrument modes, number of transits, etc.) for exoplanets of interest. This will be particularly important for smaller worlds like super-Earths, since these typically require a large number of transits to obtain the precision needed to tightly constrain atmospheric abundances. Additionally, our results imply utility for population-level studies (\textit{e.g,} via a \emph{TESS}-like mission).

These results are based on simulated spectra with no added noise, and the next step in evaluating the utility of this method is to determine its accuracy as a function of measurement precision, to assess the associated observation requirements. Incorporating instrument noise may shift effective filters, or we may need additional filters to overcome noise.

\startlongtable
\begin{deluxetable}{c|c|cc}
\tablecaption{Spectroscopic Performance for {\jwst} \& \emph{HST} Instruments for All Models and Cloud-free Models for Difference Metric\label{tab:jwstComp}}
\tablehead{\colhead{Instrument} & \colhead{Wavelengths} & \multicolumn{2}{c}{Accuracy (\%)} \\
\colhead{Mode} &\colhead{($\mu$m)} & \colhead{Cloud-Free} & \colhead{All Models}
}
\startdata
NIRISS/SOSS & 0.6 -- 2.8 & 98.5 & 98.2\\
\hline
NIRSpec Prism & 0.6 -- 5.0 & 99.2 & 99.2\\
NIRSpec G235 & 1.7 -- 3.0 & 94.7 & 95.5\\
NIRSpec G395 & 2.9 -- 5.0 & 95.4 & 95.7\\
 \hline
MIRI/LRS & 5.0 -- 13.0 & 97.1 & 97.1\\
 \hline
HST/WFC3 & 1.1 -- 1.7 & 92.5 & 93.1\\
\enddata
\end{deluxetable}

\section{Future Work} \label{sec:future}

Future papers in this series will seek to address the following enhancements for this methodology.

\subsection{Noise and Uncertainties}

Though we found that, in principle, we can differentiate between super-Earths and sub-Neptunes using a few specific filters, this analysis did not account for the effects of instrument noise, which may affect the most useful filters, or impact the general performance and practicality of this method. It will therefore be important to evaluate the performance of this method when accounting for instrument noise and uncertainties in stellar and planetary parameters.

\subsection{Clouds vs No Clouds}

We have demonstrated that transmission color analysis can be used to distinguish between low- and high-MMW atmospheres, regardless of the cloudiness of the atmosphere.  However, depending on the level of noise, clouds and high MMWs can have very similar (possibly even degenerate) muting effects on an atmosphere's transmission features, and there is value in the ability to distinguish between atmospheres with high MMWs and those with clouds for prioritization of follow-on observations by large telescopes.  Clouds flatten transmission features, degrading the information that can be gleaned from a transmission spectrum, while high-MMW atmospheres, particularly around terrestrial-sized planets, have a greater potential of containing habitability and biosignature indicators.  Therefore, knowing \emph{a priori} the likelihood of an atmosphere being cloudy will allow large telescopes to prioritize observations for planets for which we can expect to collect useful spectra.

\subsection{More MMW Groups and Expanded Multivariate Analysis}

Given the high prediction accuracies of this method for two MMW groups, we would like to expand our evaluation of this method for use in distinguishing between more MMW groups, \textit{i.e.,} more types of atmospheres (\textit{e.g.,} sub-Neptune, water-dominated, or terrestrial planets).  It is possible that differentiating between more types of atmospheres will require a different multivariate classification scheme, such as Classification and Regression Trees (CART), which, unlike LDA, allows for classification based on more than one variable (\textit{e.g.,} MMW and temperature).  If this method can be shown to categorize more types of planets with reasonably high accuracy, its utility will be greatly expanded for informing future observations and missions.

\subsection{Validation Against Observational Data}

If the transmission color analysis method can be shown to perform through the effects of instrument noise, we would next want to validate this method against data for known super-Earths and sub-Neptunes to confirm its performance against real observational data.  These planets would preferably have available spectral data in or near the wave bands discussed in Section \ref{sec:results}.  We would then construct a database of the transmission difference calculations for the corresponding filters and, as described in Section \ref{sec:metric}, run this database against the LDA from the full model set (without the 75\%\slash 25\% training\slash test split).  In this test, a high prediction accuracy would serve as validation not only of the analysis method described herein, but also of the modeling methods used to generate the simulated spectra.  A low prediction accuracy, on the other hand, could implicate either the modeling or analysis methods, or both.

\subsection{Star Type \& Atmospheric Chemistry}

This study made use of and expanded upon the atmosphere model set described in \cite{Goyal2019}.  As noted in Section \ref{sec:simspec}, these models assume a planet orbiting a Sun-like star, though we scaled the simulated transmission spectra to 0.5 R\textsubscript{Sun} to approximate the signal for a M-dwarf planet.  While the simulated spectra can be scaled for different stellar radii and various planetary parameters, the atmospheric chemistry is still based on a solar spectral energy distribution (SED), and we would expect the NIR-dominated emissions of M dwarfs to impart distinct radiative and compositional characteristics to their planets' atmospheres \citep[{\em e.g.,} ][]{Segura2005, Leconte2015}.  In the future, this analysis should be expanded to include different star types, with corresponding spectral energy distributions, and should account for atmospheric chemistry more specific to smaller planets orbiting M dwarfs.  The model spectrum population should also be expanded to incorporate a broader array of secondary atmospheres, including non-isothermal atmospheres and atmospheres dominated by disequilibrium chemistry.

\subsection{Emission Spectra}

Given the history of the reflectance color analysis method and the heretofore promising indications of this study for transmission color analysis, we are also considering expanding this analysis to evaluate use of the color method for emission spectra, which will allow us to expand beyond transiting planets and to constrain a planet's atmospheric thermal structure and circulation pattern as well as composition.

\section{Summary \& Conclusions} \label{sec:conclusion}

Considering the promising trends identified for solar system worlds and some simulated exoplanets using the reflectance color-color analysis method (see Section \ref{sec:intro}), we have explored the utility of translating this method for transmission spectroscopy. We have analyzed a grid of {\sim}19,000 simulated transmission spectra for an array of exoplanet models with variations in temperature, atmospheric metallicity (\textit{i.e.,} mean molecular weight), C/O ratio, surface gravity, cloud effects, and scattering due to hazes for over 5,000 filter pair combinations.  Given the general community interest in studying planets at the super-Earth\slash sub-Neptune boundary, we focused on the question of whether this type of analysis could allow for differentiation between these classes of planets using low-resolution transmission spectroscopy.  

We categorized our model planets by atmospheric MMW, and using Linear Discriminant Analysis, we were able to demonstrate >90\% overall accuracy in predicting a planet's category based on a few filters in the near- and mid-infrared (Table \ref{tab:filters}). Due to the smaller number of lower-MMW (\emph{i.e.,} sub-Neptune) models compared to higher-MMW models in our data set, the category prediction accuracy was lower for these models, at 81.4\% correct classifications for MMW<15 u and 68.2\% correct classifications for MMW<10 u (sub-Neptunes per our definition in Sections \ref{sec:intro} \& \ref{sec:simspec}). For our model set and categorization scheme, the LDA achieved 100\% category prediction accuracy for models with MMW>15 u.

This study has demonstrated the potential for transmission ``color'' analysis to facilitate the characterization of exoplanet atmospheres.  If further evaluation confirms its practicality notwithstanding instrument noise considerations, this method could be used for reconnaissance observations that can inform prioritization of targets and observation planning for follow-up observation by telescopes such as {\jwst}, as well as enable population-level studies of exoplanet atmospheres that will yield insights into planetary formation and evolution in addition to observational phenomena such as the Radius Gap \citep{Fulton2017} and the Neptunian desert \citep{Mazeh2016}.

\acknowledgments

\textbf{Acknowledgements}  We would like to thank the Johns Hopkins Applied Physics Laboratory Independent Research \& Development initiative and the Johns Hopkins University Earth \& Planetary Sciences Department for their support of this effort. Jayesh M. Goyal and Nikole K. Lewis acknowledge support for this work from the {\jwst} program under NASA grant 80NSSC20K0586. We would also like to thank Sarah Peacock and Jacob Lustig-Yaeger for their expertise regarding M dwarfs and M-dwarf planets.

We would also like to thank our anonymous reviewers for their constructive comments and suggestions.

\software{ATMO \citep{Amundsen2014, Tremblin2015, Tremblin2016, Tremblin2017, Drummond2016, Goyal2018, Evans2019, Goyal2019}; \hyperlink{https://github.com/natashabatalha/colorcolor}{\texttt{colorcolor}} \citep{Batalha2018}; \hyperlink{https://github.com/ksotzen/transmission_color}{\texttt{transmissioncolor}} (this work)}

\bibliography{ms}



\end{document}